\documentclass[longauth]{aa}
\usepackage[varg]{txfonts}
\bibpunct{(}{)}{;}{a}{}{,} 

\usepackage{amsmath}
\usepackage{amsfonts}
\usepackage{amssymb}
\usepackage{array}

\usepackage{natbib,twoopt}
\usepackage[breaklinks=true]{hyperref} 
\bibpunct{(}{)}{;}{a}{}{,} 

\begin{document}

\title{Search for muon neutrinos from gamma-ray bursts with the ANTARES neutrino telescope using 2008 to 2011 data}
\titlerunning{Search for muon neutrinos from GRBs with ANTARES}

\small
\author{
S.~Adri\'an-Mart\'inez \inst{\ref{UPV}} \and
A.~Albert \inst{\ref{Colmar}} \and
I.~Al~Samarai \inst{\ref{CPPM}} \and
M.~Andr\'e \inst{\ref{UPC}} \and
M.~Anghinolfi \inst{\ref{Genova}} \and
G.~Anton \inst{\ref{Erlangen}} \and
S.~Anvar \inst{\ref{IRFU/SEDI}} \and
M.~Ardid \inst{\ref{UPV}} \and
T.~Astraatmadja \inst{\ref{NIKHEF}}  \fnmsep  \and
J.-J.~Aubert \inst{\ref{CPPM}} \and
B.~Baret \inst{\ref{APC}} \and
J.~Barrios-Marti \inst{\ref{IFIC}} \and
S.~Basa \inst{\ref{LAM}} \and
V.~Bertin \inst{\ref{CPPM}} \and
S.~Biagi \inst{\ref{Bologna}, \ref{Bologna-UNI}} \and
C.~Bigongiari \inst{\ref{IFIC}} \and
C.~Bogazzi \inst{\ref{NIKHEF}} \and
B.~Bouhou \inst{\ref{APC}} \and
M.~C.~Bouwhuis \inst{\ref{NIKHEF}} \and
J.~Brunner \inst{\ref{CPPM}} \and
J.~Busto \inst{\ref{CPPM}} \and
A.~Capone \inst{\ref{Roma}, \ref{Roma-UNI}} \and
L.~Caramete \inst{\ref{ISS}} \and
C.~C$\mathrm{\hat{a}}$rloganu \inst{\ref{Clermont-Ferrand}} \and
J.~Carr \inst{\ref{CPPM}} \and
S.~Cecchini \inst{\ref{Bologna}} \and
Z.~Charif \inst{\ref{CPPM}} \and
Ph.~Charvis \inst{\ref{GEOAZUR}} \and
T.~Chiarusi \inst{\ref{Bologna}} \and
M.~Circella \inst{\ref{Bari}} \and
F.~Classen \inst{\ref{Erlangen}} \and
R.~Coniglione \inst{\ref{LNS}} \and
L.~Core \inst{\ref{CPPM}} \and
H.~Costantini \inst{\ref{CPPM}} \and
P.~Coyle \inst{\ref{CPPM}} \and
A.~Creusot \inst{\ref{APC}} \and
C.~Curtil \inst{\ref{CPPM}} \and
G.~De~Bonis \inst{\ref{Roma}, \ref{Roma-UNI}}  \and
I.~Dekeyser \inst{\ref{COM}} \and
A.~Deschamps \inst{\ref{GEOAZUR}} \and
C.~Distefano \inst{\ref{LNS}} \and
C.~Donzaud \inst{\ref{APC}, \ref{UPS}} \and
D.~Dornic \inst{\ref{CPPM}} \and
Q.~Dorosti \inst{\ref{KVI}} \and
D.~Drouhin \inst{\ref{Colmar}} \and
A.~Dumas \inst{\ref{Clermont-Ferrand}} \and
T.~Eberl \inst{\ref{Erlangen}} \and
U.~Emanuele \inst{\ref{IFIC}} \and
A.~Enzenh\"ofer \inst{\ref{Erlangen}} \and
J.-P.~Ernenwein \inst{\ref{CPPM}} \and
S.~Escoffier \inst{\ref{CPPM}} \and
K.~Fehn \inst{\ref{Erlangen}} \and
P.~Fermani \inst{\ref{Roma}, \ref{Roma-UNI}} \and
V.~Flaminio \inst{\ref{Pisa}, \ref{Pisa-UNI}} \and
F.~Folger \inst{\ref{Erlangen}} \and
U.~Fritsch \inst{\ref{Erlangen}} \and
L.~A.~Fusco \inst{\ref{Bologna}, \ref{Bologna-UNI}} \and
S.~Galat\`a \inst{\ref{APC}} \and
P.~Gay \inst{\ref{Clermont-Ferrand}} \and
S.~Gei{\ss}els\"oder \inst{\ref{Erlangen}} \and
K.~Geyer \inst{\ref{Erlangen}} \and
G.~Giacomelli \inst{\ref{Bologna}, \ref{Bologna-UNI}} \and
V.~Giordano \inst{\ref{Catania}} \and
A.~Gleixner \inst{\ref{Erlangen}} \and
J.~P.~ G\'omez-Gonz\'alez \inst{\ref{IFIC}} \and
K.~Graf \inst{\ref{Erlangen}} \and
G.~Guillard \inst{\ref{Clermont-Ferrand}} \and
H.~van~Haren \inst{\ref{NIOZ}} \and
A.~J.~Heijboer \inst{\ref{NIKHEF}} \and
Y.~Hello \inst{\ref{GEOAZUR}} \and
J.~J. ~Hern\'andez-Rey \inst{\ref{IFIC}} \and
B.~Herold \inst{\ref{Erlangen}} \and
J.~H\"o{\ss}l \inst{\ref{Erlangen}} \and
C.~W.~James \inst{\ref{Erlangen}} \and
M.~de~Jong \inst{\ref{NIKHEF}}  \fnmsep  \and
M.~Kadler \inst{\ref{Wuerzburg}} \and
O.~Kalekin \inst{\ref{Erlangen}} \and
A.~Kappes \inst{\ref{Erlangen}}  \fnmsep    \and
U.~Katz \inst{\ref{Erlangen}} \and
P.~Kooijman \inst{\ref{NIKHEF}, \ref{UU}, \ref{UvA}} \and
A.~Kouchner \inst{\ref{APC}} \and
I.~Kreykenbohm \inst{\ref{Bamberg}} \and
V.~Kulikovskiy \inst{\ref{MSU}, \ref{Genova}} \and
R.~Lahmann \inst{\ref{Erlangen}} \and
E.~Lambard \inst{\ref{CPPM}} \and
G.~Lambard \inst{\ref{IFIC}} \and
G.~Larosa \inst{\ref{UPV}} \and
D. ~Lef\`evre \inst{\ref{COM}} \and
E.~Leonora \inst{\ref{Catania}, \ref{Catania-UNI}} \and
D.~Lo Presti \inst{\ref{Catania}, \ref{Catania-UNI}} \and
H.~Loehner \inst{\ref{KVI}} \and
S.~Loucatos \inst{\ref{IRFU/SPP}, \ref{APC}} \and
F.~Louis \inst{\ref{IRFU/SEDI}} \and
S.~Mangano \inst{\ref{IFIC}} \and
M.~Marcelin \inst{\ref{LAM}} \and
A.~Margiotta \inst{\ref{Bologna}, \ref{Bologna-UNI}} \and
J.~A.~Mart\'inez-Mora \inst{\ref{UPV}} \and
S.~Martini \inst{\ref{COM}} \and
T.~Michael \inst{\ref{NIKHEF}} \and
T.~Montaruli \inst{\ref{Bari}, \ref{WIN}} \and
M.~Morganti \inst{\ref{Pisa}}  \fnmsep   \and
C.~M\"uller \inst{\ref{Bamberg}} \and
M.~Neff \inst{\ref{Erlangen}} \and
E.~Nezri \inst{\ref{LAM}} \and
D.~Palioselitis \inst{\ref{NIKHEF}} \and
G.~E.~P\u{a}v\u{a}la\c{s} \inst{\ref{ISS}} \and
C.~Perrina \inst{\ref{Roma}, \ref{Roma-UNI}} \and
P.~Piattelli \inst{\ref{LNS}} \and
V.~Popa \inst{\ref{ISS}} \and
T.~Pradier \inst{\ref{IPHC}} \and
C.~Racca \inst{\ref{Colmar}} \and
G.~Riccobene \inst{\ref{LNS}} \and
R.~Richter \inst{\ref{Erlangen}} \and
C.~Rivi\`ere \inst{\ref{CPPM}}  \fnmsep \thanks{\scriptsize{Corresponding author. Email address: \href{mailto:criviere@cppm.in2p3.fr}{criviere@cppm.in2p3.fr} (C.~Rivi\`ere)}} \and
A.~Robert \inst{\ref{COM}} \and
K.~Roensch \inst{\ref{Erlangen}} \and
A.~Rostovtsev \inst{\ref{ITEP}} \and
D.~F.~E.~Samtleben \inst{\ref{NIKHEF}, \ref{Leiden}} \and
M.~Sanguineti \inst{\ref{Genova-UNI}} \and
J.~Schmid \inst{\ref{Erlangen}}  \fnmsep  \thanks{\scriptsize{Corresponding author. Email address: \href{mailto:julia.schmid@physik.uni-erlangen.de}{julia.schmid@physik.uni-erlangen.de} (J.~Schmid) }}  \and
J.~Schnabel \inst{\ref{Erlangen}} \and
S.~Schulte \inst{\ref{NIKHEF}} \and
F.~Sch\"ussler \inst{\ref{IRFU/SPP}} \and
T.~Seitz \inst{\ref{Erlangen}} \and
R.~Shanidze \inst{\ref{Erlangen}} \and
C.~Sieger \inst{\ref{Erlangen}} \and
F.~Simeone \inst{\ref{Roma}, \ref{Roma-UNI}} \and
A.~Spies \inst{\ref{Erlangen}} \and
M.~Spurio \inst{\ref{Bologna}, \ref{Bologna-UNI}} \and
J.~J.~M.~Steijger \inst{\ref{NIKHEF}} \and
Th.~Stolarczyk \inst{\ref{IRFU/SPP}} \and
A.~S{\'a}nchez-Losa \inst{\ref{IFIC}} \and
M.~Taiuti \inst{\ref{Genova}, \ref{Genova-UNI}} \and
C.~Tamburini \inst{\ref{COM}} \and
Y.~Tayalati \inst{\ref{LPMR}} \and
A.~Trovato \inst{\ref{LNS}} \and
B.~Vallage \inst{\ref{IRFU/SPP}} \and
C.~Vall\'ee \inst{\ref{CPPM}} \and
V.~Van~Elewyck \inst{\ref{APC}} \and
P.~Vernin \inst{\ref{IRFU/SPP}} \and
E.~Visser \inst{\ref{NIKHEF}} \and
S.~Wagner \inst{\ref{Erlangen}} \and
J.~Wilms \inst{\ref{Bamberg}} \and
E.~de~Wolf \inst{\ref{NIKHEF}, \ref{UvA}} \and
K.~Yatkin \inst{\ref{CPPM}} \and
H.~Yepes \inst{\ref{IFIC}} \and
J.~D.~Zornoza \inst{\ref{IFIC}} \and
J.~Z\'u\~{n}iga \inst{\ref{IFIC}} \and
P.~Baerwald \inst{\ref{Wuerzburg}}
}
\institute{
\scriptsize{Institut d'Investigaci\'o per a la Gesti\'o Integrada de les Zones Costaneres (IGIC) - Universitat Polit\`ecnica de Val\`encia. C/  Paranimf 1, 46730 Gandia, Spain.} \label{UPV} \and
\scriptsize{GRPHE - Institut universitaire de technologie de Colmar, 34 rue du Grillenbreit BP 50568 - 68008 Colmar, France } \label{Colmar} \and
\scriptsize{CPPM, Aix-Marseille Universit\'e, CNRS/IN2P3, Marseille, France} \label{CPPM} \and
\scriptsize{Technical University of Catalonia, Laboratory of Applied Bioacoustics, Rambla Exposici\'o,08800 Vilanova i la Geltr\'u,Barcelona, Spain} \label{UPC} \and
\scriptsize{INFN - Sezione di Genova, Via Dodecaneso 33, 16146 Genova, Italy} \label{Genova} \and
\scriptsize{Friedrich-Alexander-Universit\"at Erlangen-N\"urnberg, Erlangen Centre for Astroparticle Physics, Erwin-Rommel-Str. 1, 91058 Erlangen, Germany} \label{Erlangen} \and
\scriptsize{Direction des Sciences de la Mati\`ere - Institut de recherche sur les lois fondamentales de l'Univers - Service d'Electronique des D\'etecteurs et d'Informatique, CEA Saclay, 91191 Gif-sur-Yvette Cedex, France} \label{IRFU/SEDI} \and
\scriptsize{Nikhef, Science Park,  Amsterdam, The Netherlands} \label{NIKHEF} \and
\scriptsize{APC, Universit\'e Paris Diderot, CNRS/IN2P3, CEA/IRFU, Observatoire de Paris, Sorbonne Paris Cit\'e, 75205 Paris, France} \label{APC} \and
\scriptsize{IFIC - Instituto de F\'isica Corpuscular, Edificios Investigaci\'on de Paterna, CSIC - Universitat de Val\`encia, Apdo. de Correos 22085, 46071 Valencia, Spain} \label{IFIC} \and
\scriptsize{LAM - Laboratoire d'Astrophysique de Marseille, P\^ole de l'\'Etoile Site de Ch\^ateau-Gombert, rue Fr\'ed\'eric Joliot-Curie 38,  13388 Marseille Cedex 13, France } \label{LAM} \and
\scriptsize{INFN - Sezione di Bologna, Viale Berti-Pichat 6/2, 40127 Bologna, Italy} \label{Bologna} \and
\scriptsize{Dipartimento di Fisica dell'Universit\`a, Viale Berti Pichat 6/2, 40127 Bologna, Italy} \label{Bologna-UNI} \and
\scriptsize{INFN -Sezione di Roma, P.le Aldo Moro 2, 00185 Roma, Italy} \label{Roma} \and
\scriptsize{Dipartimento di Fisica dell'Universit\`a La Sapienza, P.le Aldo Moro 2, 00185 Roma, Italy} \label{Roma-UNI} \and
\scriptsize{Institute for Space Sciences, R-77125 Bucharest, M\u{a}gurele, Romania     } \label{ISS} \and
\scriptsize{Clermont Universit\'e, Universit\'e Blaise Pascal, CNRS/IN2P3, Laboratoire de Physique Corpusculaire, BP 10448, 63000 Clermont-Ferrand, France} \label{Clermont-Ferrand} \and
\scriptsize{G\'eoazur, Universit\'e Nice Sophia-Antipolis, CNRS/INSU, IRD, Observatoire de la C\^ote d'Azur, Sophia Antipolis, France } \label{GEOAZUR} \and
\scriptsize{INFN - Sezione di Bari, Via E. Orabona 4, 70126 Bari, Italy} \label{Bari} \and
\scriptsize{INFN - Laboratori Nazionali del Sud (LNS), Via S. Sofia 62, 95123 Catania, Italy} \label{LNS} \and
\scriptsize{Mediterranean Institute of Oceanography (MIO), Aix-Marseille University, 13288, Marseille, Cedex 9, France; Universit\'e du Sud Toulon-Var, 83957, La Garde Cedex, France CNRS-INSU/IRD UM 110} \label{COM} \and
\scriptsize{Universit\'e Paris-Sud, 91405 Orsay Cedex, France} \label{UPS} \and
\scriptsize{Kernfysisch Versneller Instituut (KVI), University of Groningen, Zernikelaan 25, 9747 AA Groningen, The Netherlands} \label{KVI} \and
\scriptsize{INFN - Sezione di Pisa, Largo B. Pontecorvo 3, 56127 Pisa, Italy} \label{Pisa} \and
\scriptsize{Dipartimento di Fisica dell'Universit\`a, Largo B. Pontecorvo 3, 56127 Pisa, Italy} \label{Pisa-UNI} \and
\scriptsize{Royal Netherlands Institute for Sea Research (NIOZ), Landsdiep 4,1797 SZ 't Horntje (Texel), The Netherlands} \label{NIOZ} \and
\scriptsize{Institut f\"ur Theoretische Physik und Astrophysik, Universit\"at W\"urzburg, Am Hubland, 97074 W\"urzburg, Germany}  \label{Wuerzburg} \and
\scriptsize{Universiteit Utrecht, Faculteit Betawetenschappen, Princetonplein 5, 3584 CC Utrecht, The Netherlands} \label{UU} \and
\scriptsize{Universiteit van Amsterdam, Instituut voor Hoge-Energie Fysica, Science Park 105, 1098 XG Amsterdam, The Netherlands} \label{UvA} \and
\scriptsize{Dr. Remeis-Sternwarte and ECAP, Universit\"at Erlangen-N\"urnberg,  Sternwartstr. 7, 96049 Bamberg, Germany} \label{Bamberg} \and
\scriptsize{Moscow State University, Skobeltsyn Institute of Nuclear Physics, Leninskie gory, 119991 Moscow, Russia} \label{MSU} \and
\scriptsize{INFN - Sezione di Catania, Viale Andrea Doria 6, 95125 Catania, Italy} \label{Catania} \and
\scriptsize{Dipartimento di Fisica ed Astronomia dell'Universit\`a, Viale Andrea Doria 6, 95125 Catania, Italy} \label{Catania-UNI} \and
\scriptsize{Direction des Sciences de la Mati\`ere - Institut de recherche sur les lois fondamentales de l'Univers - Service de Physique des Particules, CEA Saclay, 91191 Gif-sur-Yvette Cedex, France} \label{IRFU/SPP} \and
\scriptsize{D\'epartement de Physique Nucl\'eaire et Corpusculaire, Universit\'e de Gen\`eve, 1211, Geneva, Switzerland} \label{WIN} \and
\scriptsize{IPHC-Institut Pluridisciplinaire Hubert Curien - Universit\'e de Strasbourg et CNRS/IN2P3  23 rue du Loess, BP 28,  67037 Strasbourg Cedex 2, France} \label{IPHC} \and
\scriptsize{ITEP - Institute for Theoretical and Experimental Physics, B. Cheremushkinskaya 25, 117218 Moscow, Russia} \label{ITEP} \and
\scriptsize{Universiteit Leiden, Leids Instituut voor Onderzoek in Natuurkunde, 2333 CA Leiden, The Netherlands} \label{Leiden} \and
\scriptsize{Dipartimento di Fisica dell'Universit\`a, Via Dodecaneso 33, 16146 Genova, Italy} \label{Genova-UNI} \and
\scriptsize{University Mohammed I, Laboratory of Physics of Matter and Radiations, B.P.717, Oujda 6000, Morocco} \label{LPMR} 
}

\authorrunning{The ANTARES Collaboration}

\abstract 
{}
{We search for muon neutrinos in coincidence with gamma-ray bursts with the ANTARES neutrino detector using data from the end of 2007 to 2011. 
}
{Expected neutrino fluxes were calculated for each burst individually.
The most recent numerical calculations of the spectra using the NeuCosmA code were employed, which include Monte Carlo simulations of the full underlying photohadronic interaction processes.  
The discovery probability for a selection of 296 gamma-ray bursts in the given period was optimised using an extended maximum-likelihood strategy.  
}
{No significant excess over background is found in the data, and 90\% confidence level upper limits are placed on the total expected flux according to the model.} 
{}


\maketitle

\section{Introduction}
Gamma-ray bursts (GRBs) are short and very intense flashes of high-energy gamma rays, which occur unpredictably and isotropically over the sky \citep{Meegan92a}. 
Over timescales of the order of a few seconds, they release as much energy as the Sun in its entire lifetime \citep[see][for a review]{Woosley06a}.
The doubly peaked distribution of burst durations measured by the BATSE satellite led \citet{Kouveliotou93a} to the classification of GRBs into two types \citep[see also][]{Paciesas99a}.
The sub-class of long bursts (with durations $\gtrsim \! 2\; \mathrm{s}$) has been shown to be associated with supernovae of type I b/c  \citep[see e.g.][]{Galama98a}. 
For the short bursts ($\lesssim \!  2\; \mathrm{s}$ duration) it is much harder to measure  the fast-fading afterglow emission and thereby obtain information about their origin. 
However, they are now widely accepted to originate from the merging of two compact objects, for example neutron stars and black holes \citep[][]{Eichler89a,Nakar07a}.

In the fireball model \citep[as proposed by][]{Meszaros93a}, the observed electromagnetic radiation is explained by highly relativistic outflows of material, most likely collimated in jets pointed towards the Earth.
Shock fronts emerge in these outflows in which electrons are accelerated \citep{Rees92a}.  
The synchrotron emission of these relativistic electrons and subsequent inverse Compton-scattering of the emitted photon field causes the observed gamma-ray radiation \citep{Meszaros06a}.
Within the framework of the fireball model, protons can also be shock-accelerated, which yields emission of high-energy neutrinos accompanying the electromagnetic signal of the burst \citep{Waxman95a,Waxman97a,Waxman00a}.
The detection of neutrinos from gamma-ray bursts would be unambiguous proof for hadronic acceleration in cosmic sources, and could also serve to explain the origin of the cosmic-ray flux at ultra-high energies \citep{Waxman95b}. 
 
Several limits over a wide range of energies have been placed on the neutrino emission from gamma-ray bursts, for instance from experiments such as Super-Kamiokande \citep{Super_Kamiokande00a}, AMANDA \citep{Amanda08a},  Baikal \citep{Baikal11a}, RICE \citep{Rice07a}, and ANITA \citep{Anita11a}. 
Very-large-volume neutrino telescopes such as ANTARES and IceCube are sensitive to neutrino fluxes above approximately $100\; \mathrm{GeV}$; 
since they simultaneously observe at least half the sky, they are ideal instruments to search for any high-energy neutrino flux from GRBs.
The data from ANTARES in its construction phase in 2007 with the first five detection lines deployed \citep{Antares13a} and from IceCube in its IC22, IC40, and IC59 detector phases from 2007 to 2010 \citep{IceCube10a,IceCube11b,IceCube12a} have previously been analysed in searches for this emission, with corresponding limits set in the TeV to PeV energy range.
 Due to its location at the South Pole, the IceCube detector has its maximum sensitivity for Northern Hemisphere sources, whereas the ANTARES detector, being situated at a latitude of $43\ensuremath{^\circ}$, has its maximum sensitivity for Southern Hemisphere sources. 

In this paper we present a search for prompt GRB neutrino emission in the period from the end of 2007 to 2011 with the ANTARES telescope using 296 bursts, out of which 90\% have not been included in the previously mentioned references.
In contrast to previous analyses, this search has for the first time been optimised for a fully numerical neutrino-emission model, based on \citet{Huemmer10a,Huemmer12a}.

This paper is organised as follows: 
in Sect.~\ref{sec:antares_detector}, the ANTARES neutrino telescope is briefly introduced. 
A description of  how the burst parameters are obtained 
and how the final sample is selected is given in Sect.~\ref{sec:grb_parameters}.
Models for the expected neutrino spectra within the fireball paradigm are described in Sect.~\ref{sec:spectra}, including the numerical model NeuCosmA, which is used for the optimisation of the analysis. 
In Sect.~\ref{sec:simulation}, we describe Monte Carlo simulations of GRB neutrino events that yield the detector response to the signal. 
The background distribution is estimated from data, as reported in Sect.~\ref{sec:background}. 
From these distributions, pseudo-experiments were generated that, by exploiting an extended maximum-likelihood ratio method, were used to optimise the search to obtain the highest discovery potential for the neutrino flux --
this is presented in Sect.~\ref{sec:pe_likelihood}. 
In Sect.~\ref{sec:optimization}, we investigate whether the discovery potential can be improved by limiting the analysis to an optimised sub-sample of the bursts.
Results of the analysis and the derived limits are provided in Sect.~\ref{sec:results}.

\section{ANTARES detector and data taking}
\label{sec:antares_detector}
The underwater neutrino telescope ANTARES \citep[see][]{Antares11a} is primarily designed for detecting relativistic muons from charged-current interactions of cosmic muon neutrinos\footnote{Throughout the paper, `neutrino' will denote both $\nu$ and $\bar{\nu}$, and `muon' will denote both $\mu^-$ and $\mu^+$.} with matter in or close-by to the detector.
The passage of these muons through the seawater induces the emission of Cherenkov light that is then detected by an array of photo-multiplier tubes (PMTs).
Using the time and position information of the photons, the muon trajectory is reconstructed. 
The original neutrino direction is then inferred from the measured muon direction -- at the energies considered here, the uncertainty introduced by the scattering angle is negligible compared to the detector's resolution.

The ANTARES telescope is located in the Mediterranean Sea at a depth of $2.4\; \mathrm{km}$. 
The detector consists of twelve vertical `strings' anchored to the seabed, each of which is held upright by a buoy at the top. They are separated from each other by a typical distance of $70\; \mathrm{m}$. The twelve strings, each with a length of $450\; \mathrm{m}$, are equipped with 25 triplets of PMTs, building a three-dimensional array of 885 PMTs in total\footnote{One string is equipped with 60 instead of 75 PMTs.}. The triplets have a vertical spacing of $14.5\; \mathrm{m}$ between them, whereas the first triplet is placed at a height of $100\; \mathrm{m}$ above the seabed. 

Bioluminescence and radioactive decay of $\element[][40]{K}$ produce a random optical background that can vary between 50 and $300\; \mathrm{kHz}$ per PMT, depending for example on the time of the year or the sea current. 
A multi-level online triggering procedure is applied to select possible particle signatures -- see \citet{Antares07a} for a more detailed description.

In addition to the cosmic neutrino signal that the ANTARES experiment is searching for, there are other processes that can produce muon tracks in the detector that are considered as background events.
Air showers are generated when high-energy cosmic rays hit the Earth's atmosphere. 
Among other particles, muons and neutrinos are produced in these showers.
Since only the (weakly interacting) neutrinos are capable of traversing the Earth, it can effectively be used as a `shield' against all particles but neutrinos. 
By searching only for upgoing particles therefore, the atmospheric downgoing muon background can be rejected. 
Nevertheless, muons from above can also produce signals in the detector that appear as upgoing events. 
Using quality cuts on the reconstruction parameters, these falsely reconstructed atmospheric muon tracks can be suppressed to a rate of $0.4$ events per day.

Atmospheric neutrinos produced by cosmic rays below the horizon can also traverse the Earth, and represent the main background component \citep[$\sim$ three events per day after quality cuts, see][]{Antares12c} to the cosmic neutrinos.

\begin{figure}[h!]
  \centering
  \includegraphics[width=\hsize]{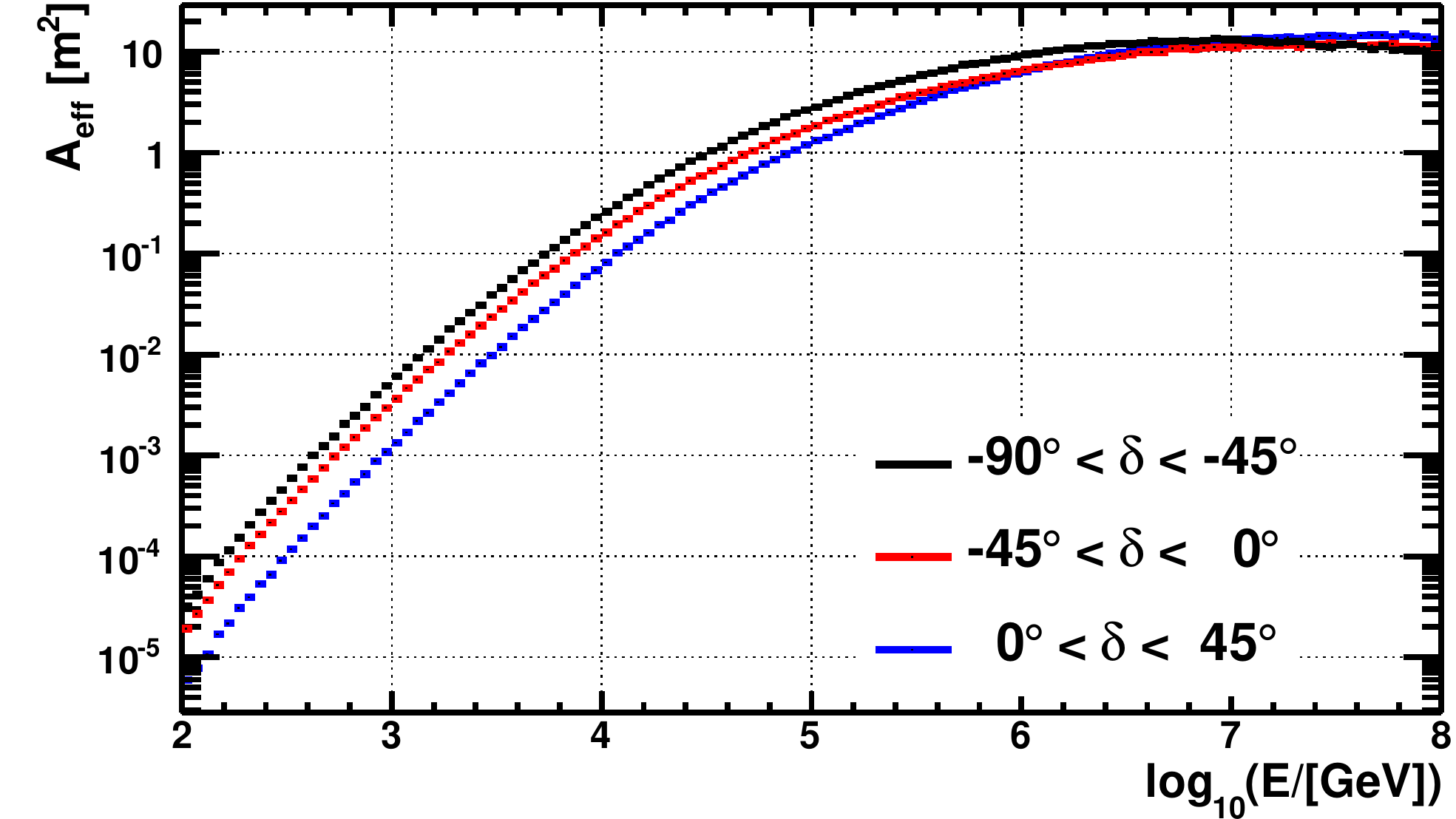}
  \caption{Time-averaged muon-neutrino effective area of the ANTARES neutrino telescope as a function of energy for different declination bands $\delta$ for the considered data-taking period. Typical quality cuts as derived in this analysis ($\Lambda>-5.35$, $\beta<1\ensuremath{^\circ}$) are applied.}
  \label{fig:eff_area}
\end{figure}

 In the analysis presented in the following, the requirement of temporal and spatial coincidence with a recorded GRB reduces the number of expected background events to $\sim \mathcal{O}(10^{-4})$ per GRB (see Sect.~\ref{sec:background}), and an extended likelihood method is furthermore used to distinguish between signal and background events.
Data collected between December 6, 2007 and the end of 2011 were analysed. 
The first six months of this period comprised the last phase of construction of the apparatus, after the deployment of detection lines 6 to 10, with the last two lines installed in May 2008. 
In that period, the instrumented volume of the detector increased from $0.008$ to $0.011\; \mathrm{km}^3$ at full size.
The corresponding average effective area to muon neutrinos as a function of the energy is shown in Fig.~\ref{fig:eff_area} for different declination bands.
The huge increase of effective area with energy, a common feature of neutrino telescopes, is due not only to the increase of the neutrino cross-section, but also to that of the muon range, which can reach several kilometres at the highest energies.
The total integrated livetime of the data in coincidence with the selected 296 GRB search-time windows is 6.6 hours.

\section{GRB selection and parameters}
\label{sec:grb_parameters}
The GRB parameters needed for the search and the simulation of expected neutrino fluxes are primarily obtained from different tables provided by the \textsl{Swift} \citep{Gehrels04a} and \textsl{Fermi} \citep{Atwood09a,Meegan09a} collaborations.
This information is then supplemented using a table supplied by the IceCube Collaboration \citep{IceCube11a}, which is created by parsing the Gamma-ray Coordinates Network (GCN) notices\footnote{GCN: \url{http://gcn.gsfc.nasa.gov/gcn3_archive.html}}.
In Appendix \ref{app:grb_selection}, we specify how these tables are merged, how often burst parameters are taken from each of them, and how the search-time windows are defined.

\begin{figure}[h!]
  \centering
  \includegraphics[width=\hsize]{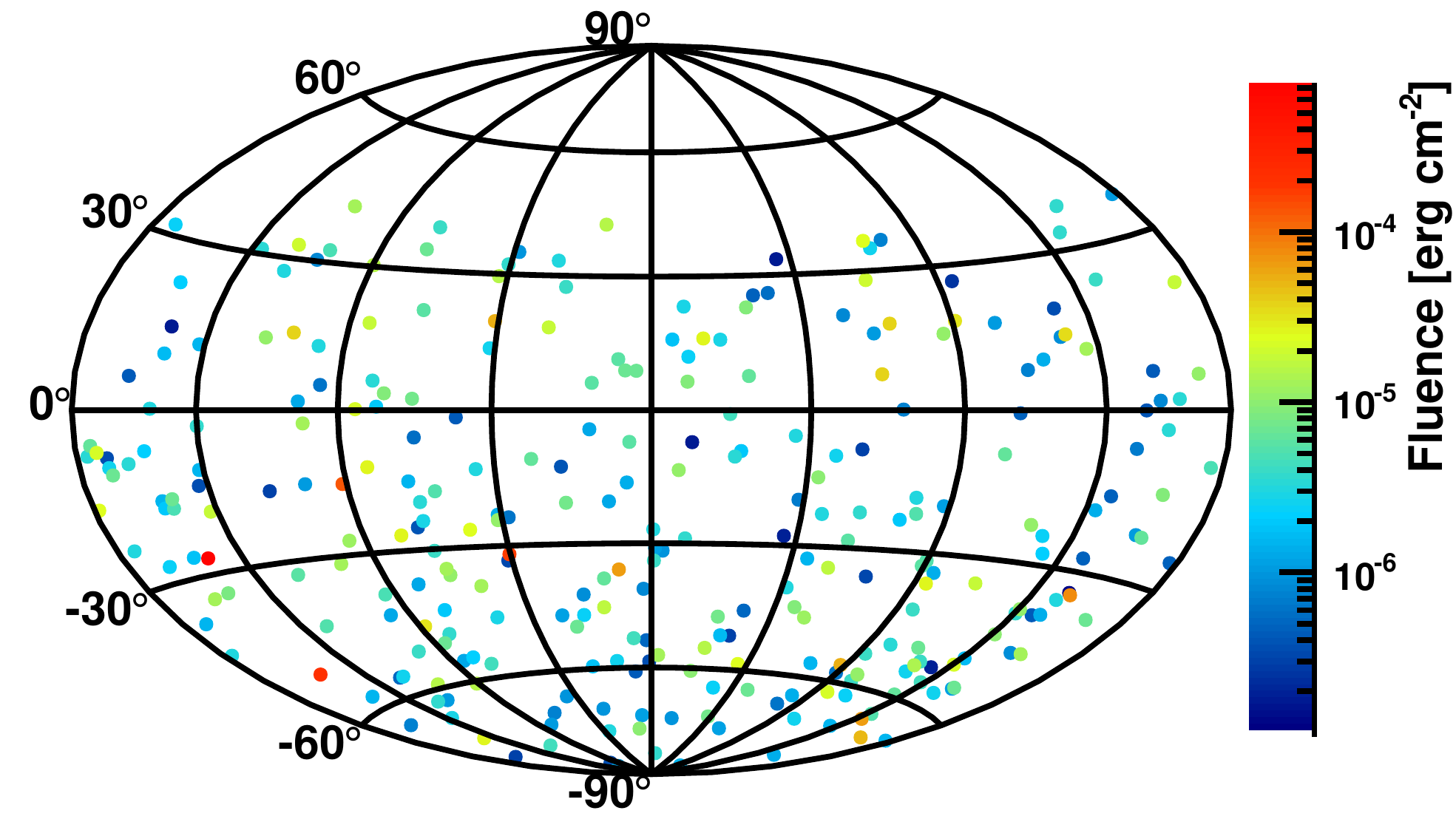}
  \caption{Sky distribution of the selected 296 GRBs in equatorial coordinates. The gamma-ray fluence of each burst is colour-coded. 
      The instantaneous field of view of the ANTARES detector is $2\pi\; \mathrm{sr}$; within a period of 24 hours, the sky up to a declination of $47\ensuremath{^\circ}$ is visible.}
  \label{fig:grb_coordinates}
\end{figure}

For the final sample, gamma-ray bursts are required to meet certain criteria as specified in Table~\ref{tab:selection} -- short bursts, for instance, are excluded as this class is much less understood.
A total of 296 bursts pass these selection cuts, of which
10\% are also included in the most recent gamma-ray-burst search from IceCube \citep{IceCube12a}.
The distribution of the selected bursts in equatorial coordinates is shown in Fig.~\ref{fig:grb_coordinates}.
\begin{table}
\caption{Selection of gamma-ray bursts} 
\label{tab:selection}  
\centering             
\begin{tabular}{l r }       
\hline\hline           
Criterion & Selected  \\
\hline                  
all GRBs (end of 2007 -- 2011)	 &  1110  \\
long GRBs 	 &  942  \\
measured spectrum &  930 \\
below ANTARES horizon 	 &  508  \\
detector running and stable data-taking conditions	 &  296   \\
\hline   
\end{tabular}
\end{table}
Out of this selection, GRB110918 outshines all others by at least half an order of magnitude in the expected neutrino flux (see Sect.~\ref{sec:spectra}). 
It is at the same time one of the most intense bursts ever observed by the \textsl{Konus-Wind} instrument \citep{Konus_Wind95a,GCN12362}.  
Unfortunately, both \textsl{Swift} and \textsl{Fermi} satellites were Earth-occulted at the time of the burst \citep{GCN110918}, but \textsl{Swift} could still observe the afterglow emission after $\sim 30\; \mathrm{h}$.
The measured parameters for this exceptional burst are given in Table~\ref{tab:default_values}. 
\section{Calculation of neutrino spectra}
\label{sec:spectra}
To calculate the expected neutrino spectra, we focused on the recently developed fully numerical NeuCosmA model \citep{Huemmer10a,Huemmer12a}.
In addition, we also present the widely used analytical approach of \citet{Guetta04a} in the following.
\subsection{Analytic approaches}
\label{sec:guetta_spectrum}
\citet{Waxman97a} were the first to calculate the expected neutrino flux in coincidence with the electromagnetic GRB in the framework of the standard fireball internal shock model, using averaged burst parameters as measured by the \textsl{BATSE} instrument on board the \textsl{CGRO} satellite \citep{Band93a}.
Their calculation was based on the assumption of Fermi-accelerated protons in the relativistic ejecta of the burst interacting with the associated photon field to produce pions via the $\Delta$-resonance. 
The subsequent decay of charged pions and muons leads to the emission of high-energy neutrinos. 
The authors derived a doubly broken power-law spectrum for the neutrinos. 
Their model is referred to as the standard Waxman-Bahcall GRB neutrino flux, and is for instance used to set limits with the BAIKAL \citep{Baikal11a} and AMANDA \citep{Amanda08a} experiments.
\citet{Guetta04a} modified the formulae of Waxman and Bahcall to calculate individual neutrino fluxes for the bursts.
Such individual burst predictions are used in recent searches with IceCube \citep{IceCube10a,IceCube11b,IceCube12a}, RICE \citep{Rice07a}, and ANITA \citep{Anita11a}.
Note in particular that the most recent IceCube limit was a factor of 2.1 below predictions made using this model.
This could either indicate the need for rejection of the model, a modification of the parameters upon which it is based, or for more detailed modelling of the neutrino emission within the fireball paradigm. 

To calculate the analytic spectrum as shown in Fig.~\ref{fig:spectra}~(a), blue solid line, the formulae given in \citet[Appendix A]{IceCube10a} were applied. 

In principle, \citet{Guetta04a} predicted different break energies for $\nu_\mu$ and $\bar{\nu}_\mu$ \citep[see][Eqs.~A10 and A11]{Guetta04a}, yielding three breaks in the combined $\nu_\mu + \bar{\nu}_\mu$ spectrum (or in the single $\nu_\mu$ spectrum when taking oscillations into account) as shown in Fig.~\ref{fig:spectra}~(a), blue dashed -- In the previous ANTARES analysis \citep{Antares13a}, this effect has been accounted for.

\subsection{NeuCosmA model}
\label{sec:neucosma_spectrum}
Monte Carlo algorithms such as SOPHIA \citep[see][]{Mucke00a} have improved the calculations of photohadronic interactions in astrophysical processes. 
\citet{Huemmer12a} showed that a more detailed treatment of the particle physics involved in calculating the neutrino spectra greatly changes the resulting neutrino flux predictions compared with those described in Sect.~\ref{sec:guetta_spectrum}.
Using their neutrinos from cosmic accelerators (NeuCosmA) code, the authors pointed out how the full photohadronic interaction cross-section, individual treatment of secondary particles (including energy losses), and neutrino mixing affect the predicted neutrino flux \citep[][]{Huemmer10a,Huemmer12a,Baerwald12b}. 
Owing to these effects, the doubly peaked structure of the spectrum from muon and pion decays now features an additional high-energy component from $K^+$ decays.
Moreover, the authors discussed several problems with the older neutrino flux estimations, such as using the real average photon energy instead of the peak energy of the photon distribution, taking the full width of the $\Delta$-resonance into account, and simulating the energy losses of secondary particles  as well as the energy dependence of the mean free path of protons \citep[see also][]{He12a}. 
The combination of all these effects gives rise to a prediction for the neutrino yield that is about one order of magnitude below the result of \citet{Guetta04a}.

Note that the NeuCosmA model, however, does not introduce any new assumptions on the nature of GRBs in general, but applies known physics in greater detail within the paradigm of the fireball model.
Hence, limits on these predictions will first of all constrain parameters governing the fireball model (such as the boost factor $\Gamma$ of the jet), and might later on probe the fireball paradigm itself. 

\begin{figure}[h!]
  \centering
  \includegraphics[width=\hsize]{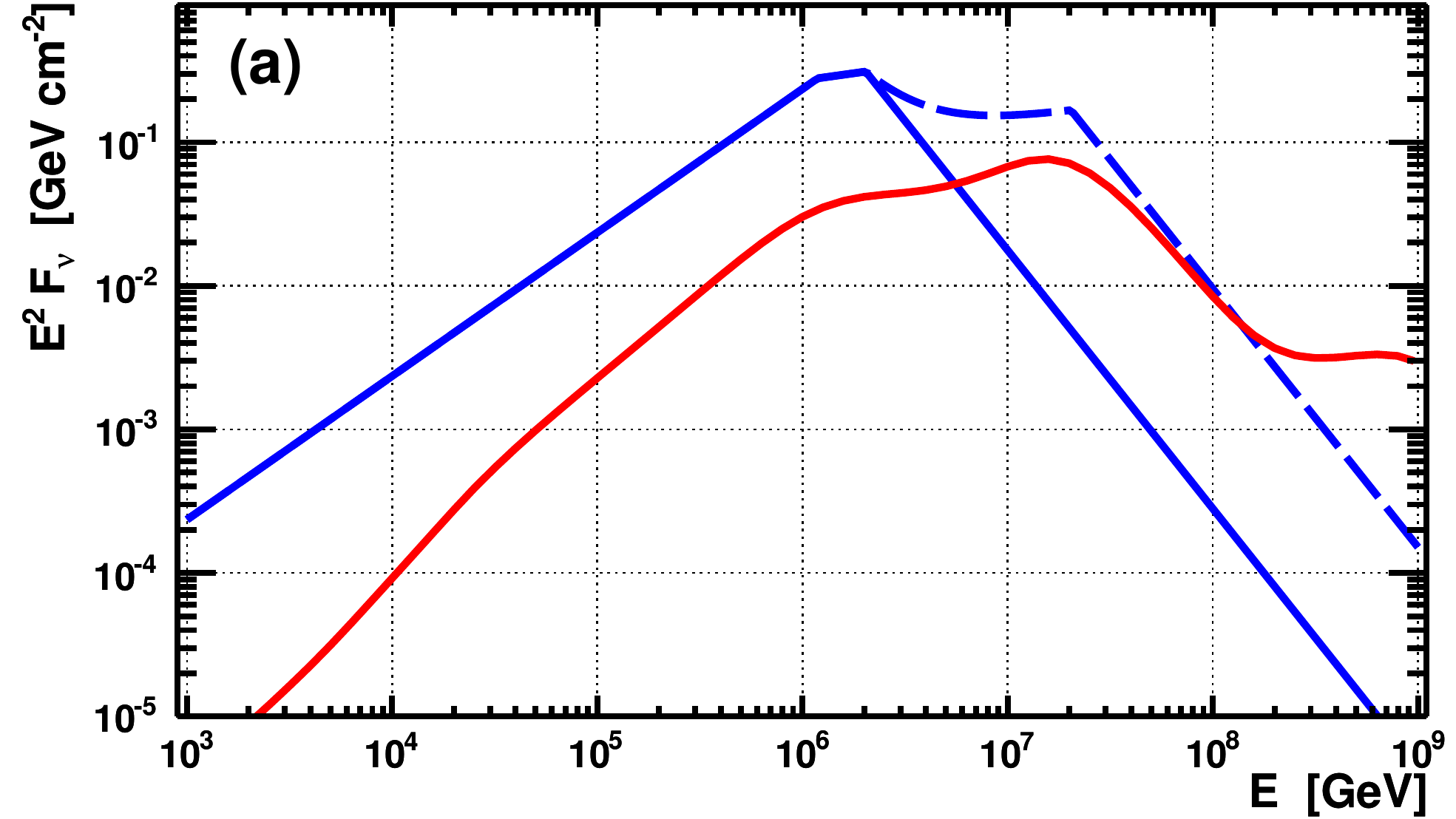}	
  \includegraphics[width=\hsize]{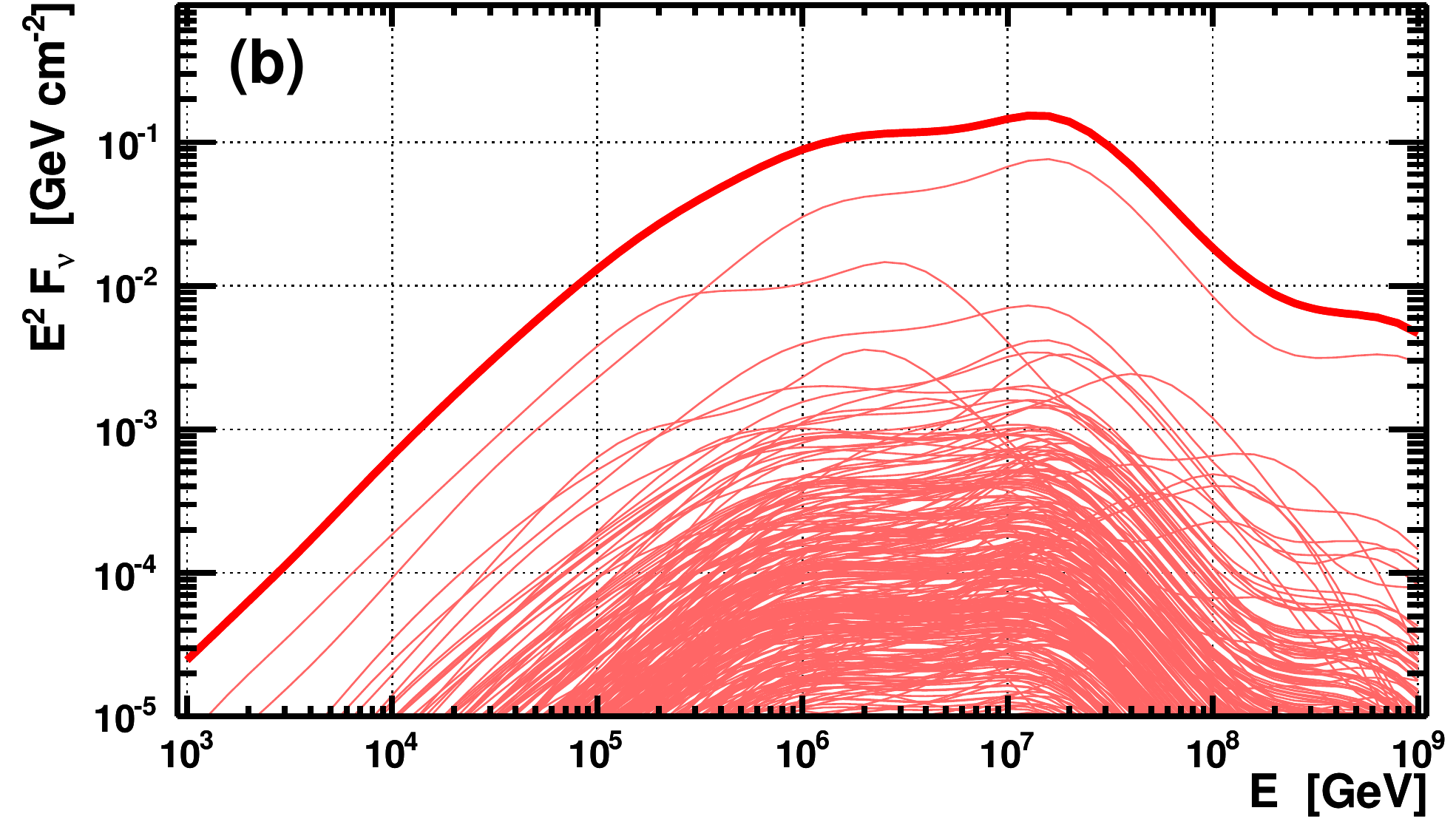}		
  \caption{\label{fig:spectra}
    (a): Expected  $\nu_\mu + \bar{\nu}_\mu$ spectra of GRB110918. The analytic model of \citet{Guetta04a} (blue) is shown with the usual simple treatment (blue solid) and accounting for different break energies of $\nu_\mu$ and $\bar{\nu}_\mu$ (blue dashed). The numerical NeuCosmA prediction is presented in red.
    (b): Individual $\nu_\mu + \bar{\nu}_\mu$ NeuCosmA spectra of the 296 GRBs (thin lines) and their sum (thick line).
  }
\end{figure}

Predictions made by the NeuCosmA model are used to optimise this analysis.
In Fig.~\ref{fig:spectra} (a), a comparison between the predicted neutrino spectra from \citet{Guetta04a} and from NeuCosmA is shown exemplarily for GRB110918. 
Fig.~\ref{fig:spectra} (b) shows all individual numerical neutrino spectra and their sum.

\section{Simulation and signal probability density function}
\label{sec:simulation} \label{sec:signal_psf}
For each gamma-ray burst in the selection, neutrino events were generated with high statistics to simulate the predicted NeuCosmA spectrum.
They were then reconstructed to compute the acceptance of the detector.
Their distribution gives the signal probability density function (PDF) labelled $\mathcal{S}(\alpha) = \mathrm{d} N (\alpha) / \mathrm{d} \Omega$, with $\alpha$ being the space angle between the reconstructed track direction and the gamma-ray-burst's coordinates. 

Firstly, the passage of neutrinos from the direction of a GRB through the Earth was simulated. 
If they interacted with matter sufficiently close to or in the detector, 
the resulting hadronic shower from the break-up of the target nucleon was generated.
The secondary particles were then propagated through the medium, inducing the emission of Cherenkov photons on their trajectory. 
The response of the PMTs to the emitted Cherenkov light was then simulated, taking into account the detector and environmental conditions at the time of the GRB occurrence. 
For a more detailed description of the simulation scheme, see \citet{Antares12c}.  

For each GRB, $4\cdot 10^{9}$ muon neutrinos were simulated.
No background events were generated, as the background rate was estimated using the data themselves (see Sect.~\ref{sec:background}).

\label{sec:reconstruction}
The applied track reconstruction algorithm is the same as used in \citet{Antares12c}.
It is based on a multi-step algorithm to maximise the likelihood of an assumed track hypothesis. 
The reconstruction returns two quality parameters, namely the track-fit quality parameter $\Lambda$ and the estimated angular uncertainty on the muon track direction $\beta$. 
Cuts on these parameters can be used to improve the signal-to-noise ratio. 
To ensure a good directional reconstruction of the selected neutrino candidates, we required $\beta < 1\ensuremath{^\circ}$. 
For a sample of track candidates with high reconstruction quality ($\Lambda > -5.2$), this cut has been shown to remove most of the remaining atmospheric muons that were falsely reconstructed as upgoing without significantly affecting the neutrino signal:
atmospheric muons are reduced by a factor of $1.2\cdot 10^{-5}$, while atmospheric neutrinos are reduced by $0.19$. 
The total background due to atmospheric events is decreased by $10^{-6}$ \citep[][]{Antares12c}.
In this analysis, the same cut combination would leave $\sim 60 \dots 70\%$ of a typical GRB signal.
However, the narrow time windows (typically a few tens of seconds, see Table~\ref{tab:analysis_params} for per-GRB values) yield intrinsically low background in coincidence with each gamma-ray burst, allowing one to loosen the cuts on the $\Lambda$ parameter. 
These were then optimised for each burst individually.

To account for the satellite's uncertainty on the direction of the GRB, the reconstructed space angle was additionally smeared with a Gaussian of the appropriate width.

\begin{figure}[h!]
  \centering
  \includegraphics[width=\hsize]{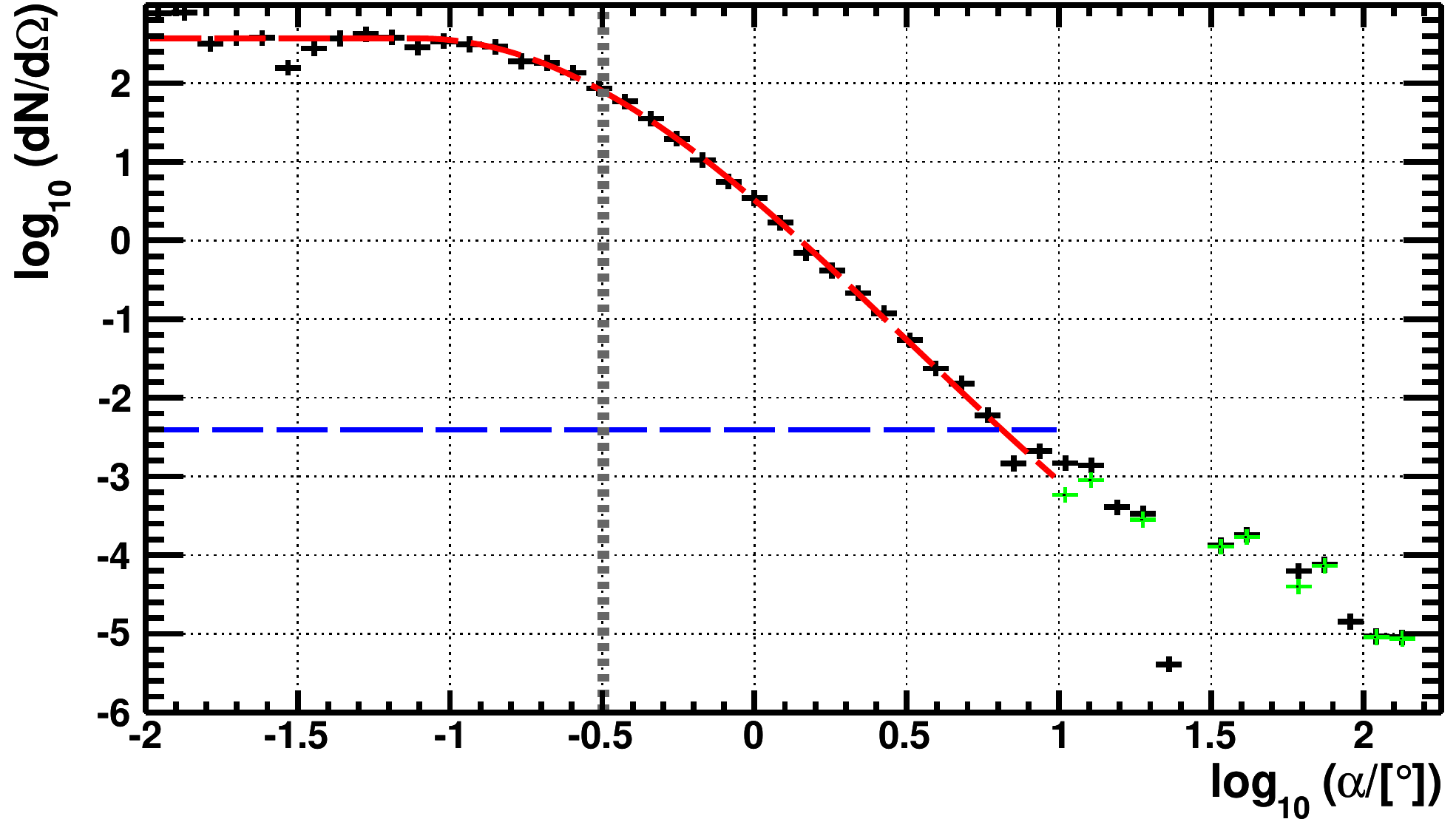}
  \caption{Simulated and reconstructed signal events per solid angle $\Omega$ versus the logarithm of the space angle $\alpha$ in degrees for the burst GRB110918: muon tracks are plotted in black, shower-like events are drawn in green. The corresponding fit is shown in red (see Eq.~\ref{eq:dNdOmega}).
    The grey dotted line indicates the median angular spread of events $\langle \alpha \rangle=0.32\ensuremath{^\circ}$. 
    The blue dashed line shows the flat background distribution $\mathcal{B}(\alpha)$ as calculated in Sect.~\ref{sec:background}. 
    Cut values $\Lambda>-5.5$ and $\beta<1\ensuremath{^\circ}$ are applied here.
  }
  \label{fig:signal_psf}
\end{figure}

The resulting distribution of events relative to the GRB direction for each cut on $\Lambda$ was then fitted with the function
\begin{align}
  \log \mathcal{S}(\alpha) = \log\frac{\mathrm{d} N (\alpha)}{\mathrm{d} \Omega}  &=
  \begin{cases}
    A, & \mbox{if }  \alpha \le  \alpha_0 \\
    A-B\cdot \left( 1-\exp\left(\frac{-(\log \alpha - \log \alpha_0)^2}{2\sigma^2}\right) \right)  & \mbox{if }  \alpha > \alpha_0 \; ,
  \end{cases}  \label{eq:dNdOmega} 
\end{align} 
with the free parameters  $A$, $B$, $\alpha_0$, and $\sigma$ as shown in Fig.~\ref{fig:signal_psf}.
In the final analysis, events with an angular distance of up to $10\ensuremath{^\circ}$ from the burst positions were included in the maximum-likelihood search.

An investigation of shower-like events (e.g. from neutral current interactions of $\nu_\mu$) revealed that these contribute only negligibly to the signal PDF up to $10\ensuremath{^\circ}$, as the original neutrino direction cannot be adequately reconstructed because of the spherical signatures of these interactions. 

\section{Background estimation}
\label{sec:background}
Upgoing atmospheric neutrinos constitute the main background component for each GRB, with a smaller contribution coming from misreconstructed downgoing atmospheric muons.
To estimate the mean number of background events $\mu_\mathrm{b}$ for each $\Lambda$ cut for each burst as realistically as possible, data were used.
However, as the number of upgoing events is very low \citep[$\sim$ four per day, see][Table 1]{Antares12c}, long time periods are needed to yield sufficient statistics, which in turn requires averaging over different data-taking conditions (in particular because of seasonal variations of the optical background). 
To compensate for this, we first estimated the average rate of reconstructed events in the GRB's direction $\langle  \mu (\theta, \phi)_\mathrm{GRB}  \rangle_\text{all runs}$ in local coordinates zenith $\theta$ and azimuth $\phi$
using data from the entire 2007 to 2011 period, then adjusted it for varying data-taking conditions (for a detailed description, see Appendix~\ref{app:background}). 

For the final search, the number of background events in coincidence with each burst and within a search cone of $10\ensuremath{^\circ}$ around its position was evaluated to be of the order of $10^{-4}$.
The background PDF, $\mathcal{B}(\alpha)$, was considered to be flat in $\Omega$ within this cone, as shown in Fig.~\ref{fig:signal_psf} for GRB110918.

\section{Pseudo-experiments and extended maximum-likelihood ratio}
\label{sec:pe_likelihood}
In the following, we show how pseudo-experiments were generated to compute the significance of a measurement, then how the cut on the reconstruction quality parameter $\Lambda$ was optimised for each GRB to yield the highest discovery probability for a signal according to the NeuCosmA model. 

For the pseudo-experiments, signal and background events $i$ with space angle $\alpha_i$ were drawn randomly from the normalised signal $\mathcal{S}(\alpha)$ and background $\mathcal{B}(\alpha)$ PDFs corresponding to each chosen cut on $\Lambda$.
For each pseudo-experiment with a total number of events $n_\mathrm{tot}$, the test statistic $Q$ was calculated as follows: 
\begin{equation}
  Q = \max_{\mu^{\prime}_\mathrm{s} \in [0,n_\mathrm{tot}]} \left( \sum\limits_{i=1}^{n_\mathrm{tot}} \log \frac{\mu^\prime_\mathrm{s} \cdot \mathcal{S}(\alpha_i) + \mu_\mathrm{b} \cdot \mathcal{B}(\alpha_i)}{\mu_\mathrm{b} \cdot \mathcal{B}(\alpha_i)}  - (\mu^{\prime}_\mathrm{s} +\mu_\mathrm{b}) \right).
  \label{eq:ex_max_likelihood}
\end{equation}
This is the so-called extended maximum-likelihood ratio \citep{Barlow90a} with an a priori knowledge of the  expected number of background events $\mu_\mathrm{b}$ (as evaluated in Sect.~\ref{sec:background} and Appendix~\ref{app:background}).
Higher values of $Q$ indicate that the measurement is more compatible with the signal hypothesis.
The signal contribution $\mu^{\prime}_\mathrm{s}$ is scanned between $0$ and $n_\mathrm{tot}$, its value corresponding to the maximum of the sum in Eq.~\ref{eq:ex_max_likelihood} returns the  estimated signal $\mu_\mathrm{s}^\text{est}$.

\begin{figure}[h!]
  \centering
  \includegraphics[width=\hsize]{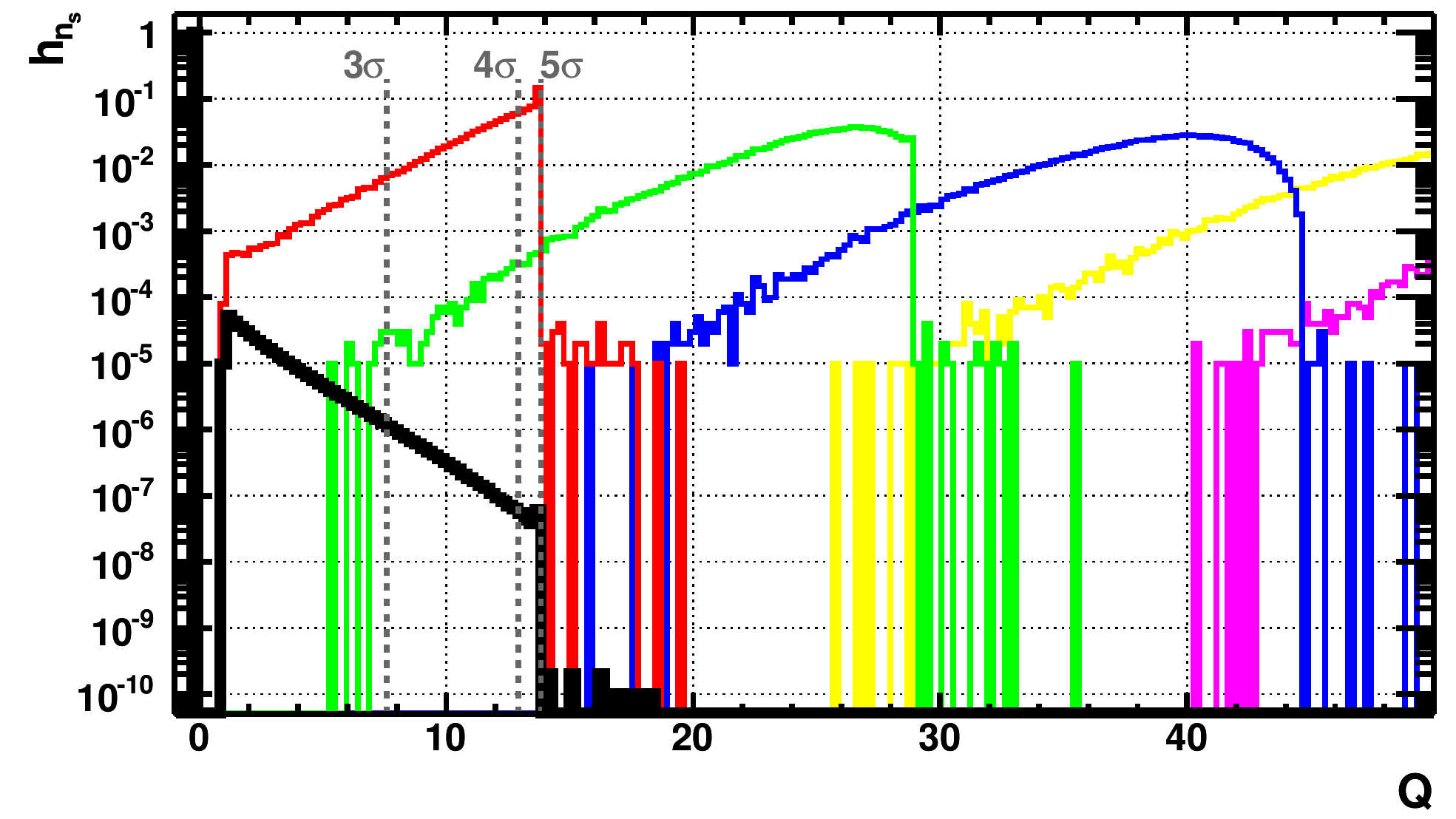}
  \caption{Probability distributions of $Q$-values, $ h_{n_\mathrm{s}}(Q)$, for different numbers  of injected signal events, $n_\mathrm{s}$. Black: background only, $ h_{0}(Q)$; red, green, blue$\dots$: $n_\mathrm{s}=1,2,3\dots$ injected signal events.
    Grey vertical lines indicate the threshold values $Q_{p}^\text{thres}$ for different significances after accounting for a trial factor of 296 (see Sect.~\ref{sec:optimization}) as calculated from $ h_{0}(Q)$.
    This example shows GRB110918 with $\Lambda > -5.5$ and $\mu_b=3.7\cdot 10^{-4}$ background events.}
  \label{fig:pe_q_hist}
\end{figure}

\begin{figure}[h!]
  \centering
  \includegraphics[width=\hsize]{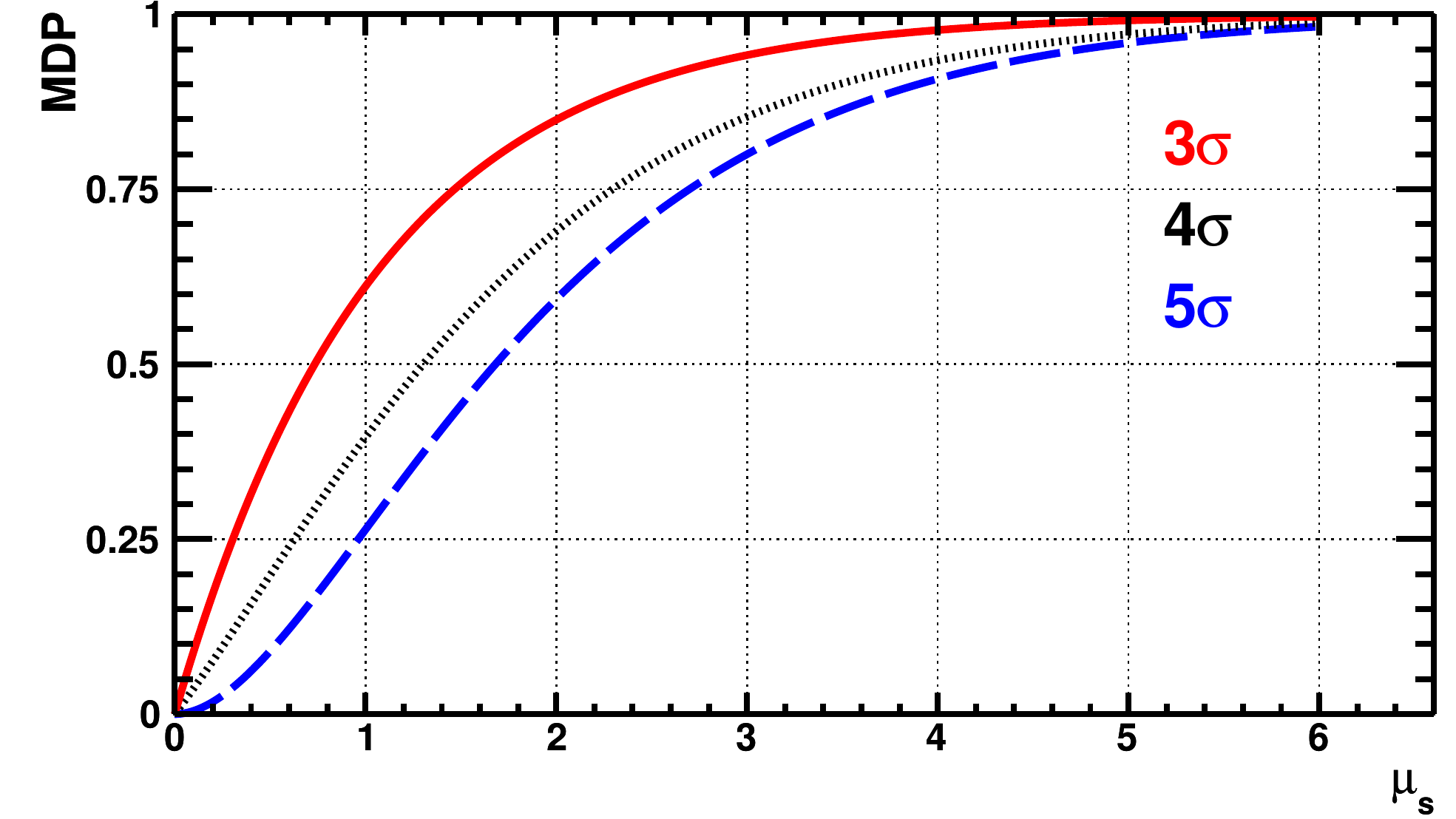}
  \caption{Model discovery potential $\mathcal{MDP}$ versus number of signal events $\mu_\mathrm{s}$ for $3\sigma$ (red solid line), $4\sigma$ (black dotted), and $5\sigma$ (blue dashed) for GRB110918.
  }
  \label{fig:pe_mdp}
\end{figure}

In the following, $h_{n_\mathrm{s}}(Q)$ denotes the distribution of $Q$-values for $n_\mathrm{s}$ injected signal events with a Poisson-distributed number of background events with expectation value $\mu_\mathrm{b}$.
The significance of a measurement is determined by its $p$-value\footnote{We used the two-sided convention, that is, $p_{3 \sigma} = 2.7 \cdot 10^{-3}$, $p_{5 \sigma} = 5.7 \cdot 10^{-7}$.}, which is given by the probability to yield $Q$-values at least as high as that observed if the background-only hypothesis were true.
Hence, using the background-only distribution $h_{0}(Q)$, the lowest $Q$-value $Q^\text{thres}_{p}$ that is necessary to claim a discovery with a certain $p$-value can be calculated with
\begin{equation}
  \begin{aligned}
    P (Q \geq Q_{p}^\text{thres} | \, \mu_\mathrm{b}) = \int_{Q_{p}^\text{thres}}^\infty  h_{0}(Q) \, \mathrm{d} Q & = p \,.
  \end{aligned}
\end{equation}
The probability distributions $h_{n_\mathrm{s}}(Q)$ for different $n_\mathrm{s}$ are shown in Fig.~\ref{fig:pe_q_hist}, with the threshold $Q$-values indicated by the grey dashed lines.
The probability distribution of $Q$ values for any number of expected signal events $\mu_\mathrm{s}$ is calculated via
\begin{equation} \label{eq:p_q_mu}
  P(Q | \, \mu_\mathrm{s}) = \sum_{n_\mathrm{s}=0}^{\infty} \mathcal{P}(n_\mathrm{s} | \, \mu_\mathrm{s}) \cdot h_{n_\mathrm{s}}(Q) \; ,
\end{equation}
with the Poisson distribution $\mathcal{P}(n_\mathrm{s} | \, \mu_\mathrm{s})$ giving the probability of observing $n_\mathrm{s}$ events from a mean number of expected events $\mu_\mathrm{s}$.
The integral of $P(Q | \, \mu_\mathrm{s}) $ gives the model discovery potential $\mathcal{MDP}$; it is the probability to make a discovery assuming that the model was correct:
\begin{eqnarray}
  \mathcal{MDP} &=&  P(Q \geq Q_{p}^\text{thres} | \, \mu_\mathrm{s}) = \int_{Q_{p}^\text{thres}}^\infty P(Q | \, \mu_\mathrm{s}) \, \mathrm{d} Q \\  \nonumber
  &=&
  \sum_{n_\mathrm{s}=0}^{\infty} \mathcal{P}(n_\mathrm{s} | \, \mu_\mathrm{s}) \cdot  \int_{Q_{p}^\text{thres}}^\infty h_{n_\mathrm{s}}(Q) \, \mathrm{d} Q.
  \label{eq:mdp}
\end{eqnarray}
The value of the $\Lambda$ cut at a given significance level is then chosen as that which maximises the $\mathcal{MDP}$ for the value of $\mu_\mathrm{s}$ predicted by the NeuCosmA model (see Sect.~\ref{sec:neucosma_spectrum}). 
Fig.~\ref{fig:pe_mdp} shows $\mathcal{MDP}(\mu_\mathrm{s})$ of GRB110918 for $3\sigma$, $4\sigma$, and $5\sigma$ versus an arbitrary number of signal events.  
The distribution $P(Q | \, \mu_\mathrm{s})$ from Eq.~\ref{eq:p_q_mu} was also used to set upper limits on the number of signal events when no discovery is made. 
For example, when the final analysis returns $Q_\mathrm{meas}$, we set a 90\% confidence level (C.L.) upper limit $\mu_\mathrm{s}^{90\%}$ on the signal by rejecting all event expectations $\mu_\mathrm{s}$ that would lead to values $Q>Q_\mathrm{meas}$ in 90\% of all pseudo-experiments:
\begin{equation} 
  P(Q \geq Q_\mathrm{meas} | \, \mu_\mathrm{s}^{90\%})  = \int_{Q_\mathrm{meas}}^\infty P(Q |\; \mu_\mathrm{s}^{90\%}) \,\mathrm{d} Q = 0.9 \, .
\end{equation}
When no event was measured ($Q_\mathrm{meas}=0$), a 90\% C.L. upper limit was set at 2.3, the lowest possible value\footnote{The value derives from Poisson statistics, since the probability to detect at least one event at a mean rate of 2.3 is exactly $90\%$.}.

For each number of injected signal events $n_\mathrm{s}$, $10^{5}$ pseudo-experiments were generated\footnote{Far more background-only pseudo-experiments are required to allow determining $Q_{p}^\text{thres}$ at $p$-values as low as $p_{5 \sigma}/296 \sim 2\cdot 10^{-9}$.} to derive the signal distributions $h_{n_\mathrm{s}}(Q)$.
Using this procedure, the model discovery potential was calculated for any given $\mu_\mathrm{s}$, and the final cut on $\Lambda$ for each GRB was found. 

\section{Search optimisation}
\label{sec:optimization}
\begin{figure}[h!] \centering
  \includegraphics[width=\hsize]{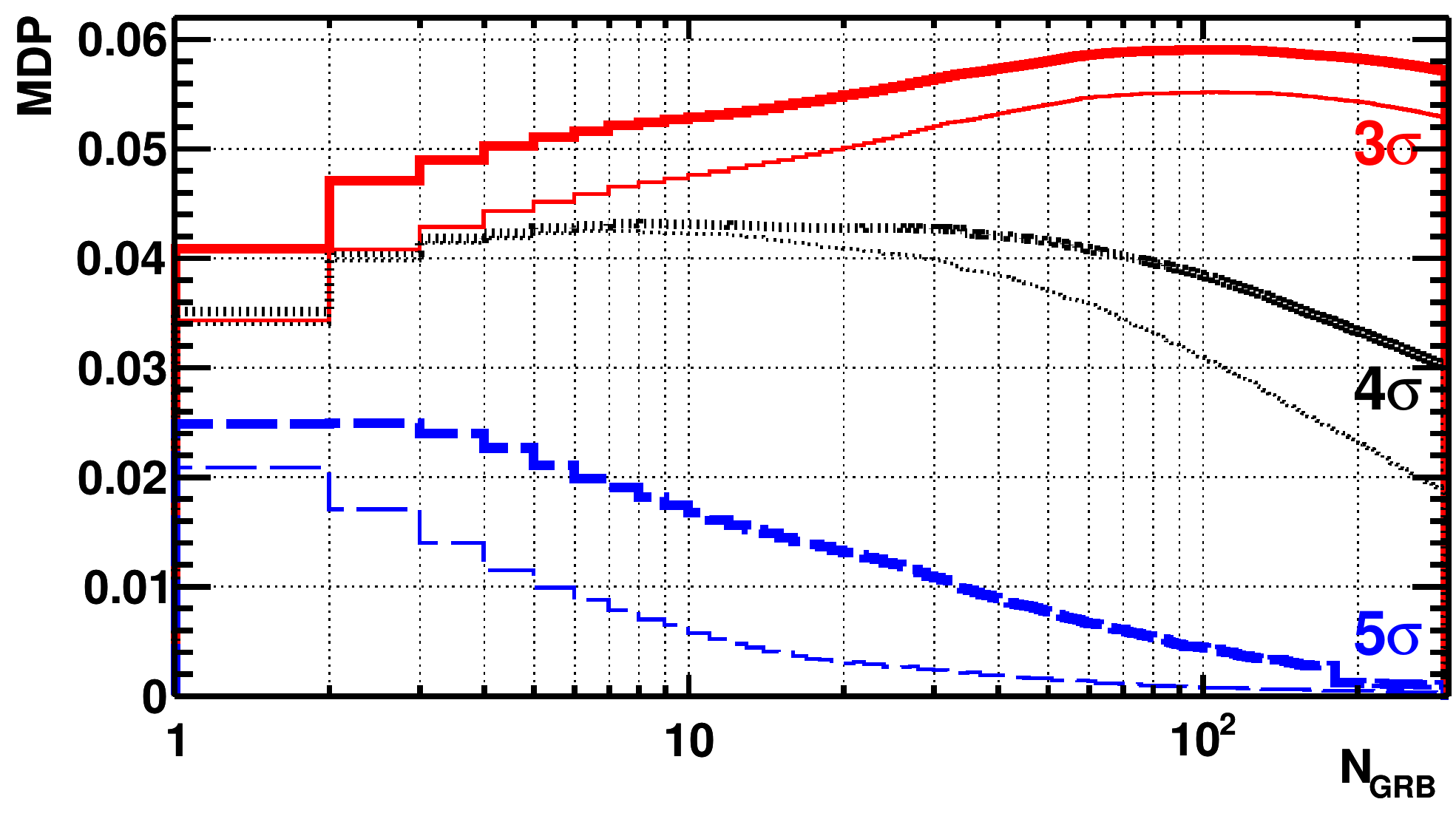}
  \caption{\label{fig:mdp_distribution} 
    Model discovery potential $\mathcal{MDP}$ versus the number of GRBs in an optimised sub-sample, $N_\mathrm{GRB}$, for $3\sigma$, $4\sigma$, and $5\sigma$ in red solid, black dotted, and blue dashed lines. For each sub-sample, only the  $N_\mathrm{GRB}$ bursts with the best $\mathcal{MDP}_i$ at the given trial factor $N_\mathrm{GRB}$ are chosen.
    The thick lines show the $\mathcal{MDP}$ distributions of the likelihood method used in this analysis, the thin lines show the distributions for a simple counting search with fixed quality cuts $\Lambda>-5.5$, $\beta<1^\circ$ (see text).
  }
\end{figure}

In the following, the best trade-off between an increased sample and the associated statistical penalty is investigated.
In general, a weighting factor, $w_i$, could be assigned to each GRB according to the predicted flux from the model.
However, this would result in the search being very sensitive to the combined uncertainty from the NeuCosmA model and especially to the parameters upon which it is based. 
An alternative approach is to include only the $N_\mathrm{GRB}$ most promising candidates, where to maintain the same overall probability of making a false discovery, the $p$-value for each burst must be divided by the total trial factor $N_\mathrm{GRB}$. 
By ordering the bursts from highest to lowest $\mathcal{MDP}_i$, the most promising $N_\mathrm{GRB}$ can be chosen to maximise the combined probability of making a significant discovery from any of them. 
The total model discovery potential is then calculated via
\begin{equation}
\mathcal{MDP}(N_\mathrm{GRB}) = 1 - \prod_{i=1}^{N_\mathrm{GRB}} (1 - \mathcal{MDP}_i).
\label{eq:mdp_n_grb}
\end{equation}
The resulting distributions of $\mathcal{MDP}(N_\mathrm{GRB})$ as a function of the size of the considered sub-sample of GRBs are shown in Fig.~\ref{fig:mdp_distribution}  for $3\sigma$, $4\sigma$ and $5\sigma$ (thick lines). 
The $3\sigma$ $\mathcal{MDP}$ distribution rises to a maximum of 0.059 at a sample size of 106 GRBs, but is relatively flat around its highest value. 
For a search optimised for $3\sigma$ it is therefore reasonable to take the whole set of GRBs, as it does not decrease the $\mathcal{MDP}$ significantly (around $3.4\%$) and the search remains less model-dependent. 
The $5\sigma$ $\mathcal{MDP}$ distribution, on the other hand, is prominently peaked at $N_\mathrm{GRB}=1$, with $N_\mathrm{GRB}=2$ being almost equivalent ($\mathcal{MDP}(1) = \mathcal{MDP}(2) =0.025$) -- the model discovery potential then decreases for larger samples. 
Even for $N_\mathrm{GRB}=2$, the second strongest GRB contributes only a small fraction to the discovery potential. 

For comparison, the distributions for a simple counting search (in which all events passing the cut criteria carry equal weighting) are also shown in Fig.~\ref{fig:mdp_distribution} (thin lines).
To mimic this search, a radius cut $\alpha_\mathrm{cut}$ for each GRB was calculated from the known background $\mu_\mathrm{b}$ at fixed reasonable quality cuts ($\Lambda>-5.5, \beta<1^\circ$) and the given significance level $p/N_\mathrm{GRB}$. 
Applying this search radius cut on the signal PDF $\mathcal{S}(\alpha)$, the expected number of signal events $\mu_\mathrm{s}$ can be estimated and consequently the $\mathcal{MDP}_i$ evaluated as the probability of detecting more than zero events.

As expected, the $\mathcal{MDP}$ curves for a counting analysis are well below those for the likelihood method,
showing the advantage of the search method used in this analysis. 
The shapes of the curves, on the other hand, are quite similar, and the same conclusions can be drawn from them, namely that using the whole sample gives the best discovery probability at $3\sigma$ and using only the individual GRB110918 at $5\sigma$.

Based on these results, we decided to optimise the quality cut on $\Lambda$ for a likelihood search on the whole sample of 296 GRBs at the $3\sigma$ significance level.
Because $3\sigma$ is not enough to claim a discovery, we predefined a cut on $\Lambda$ that was optimised for a $5\sigma$ discovery, which was then used for a separate search for the emission from GRB110918 only.

Optimised cuts $\Lambda_\mathrm{cut}$ for the final analysis as well as the accordingly expected number of background and signal events, the median angular resolution, and the search-time window are shown in Table~\ref{tab:analysis_params} for the ten most promising gamma-ray bursts. 
A full list for the 296 selected bursts can be found in the online.

\section{Results and discussion}
\label{sec:results}
Using the strategy outlined above, we analysed ANTARES data from the end of 2007 to 2011 searching for neutrino events in coincidence with the search-time windows and within $10\ensuremath{^\circ}$ around each gamma-ray burst. 
No data events passed this selection within the accumulated search duration of 6.6 hours.
Hence, the measured $Q$-value is zero. 

\begin{figure}[h!] \centering
  \includegraphics[width=\hsize]{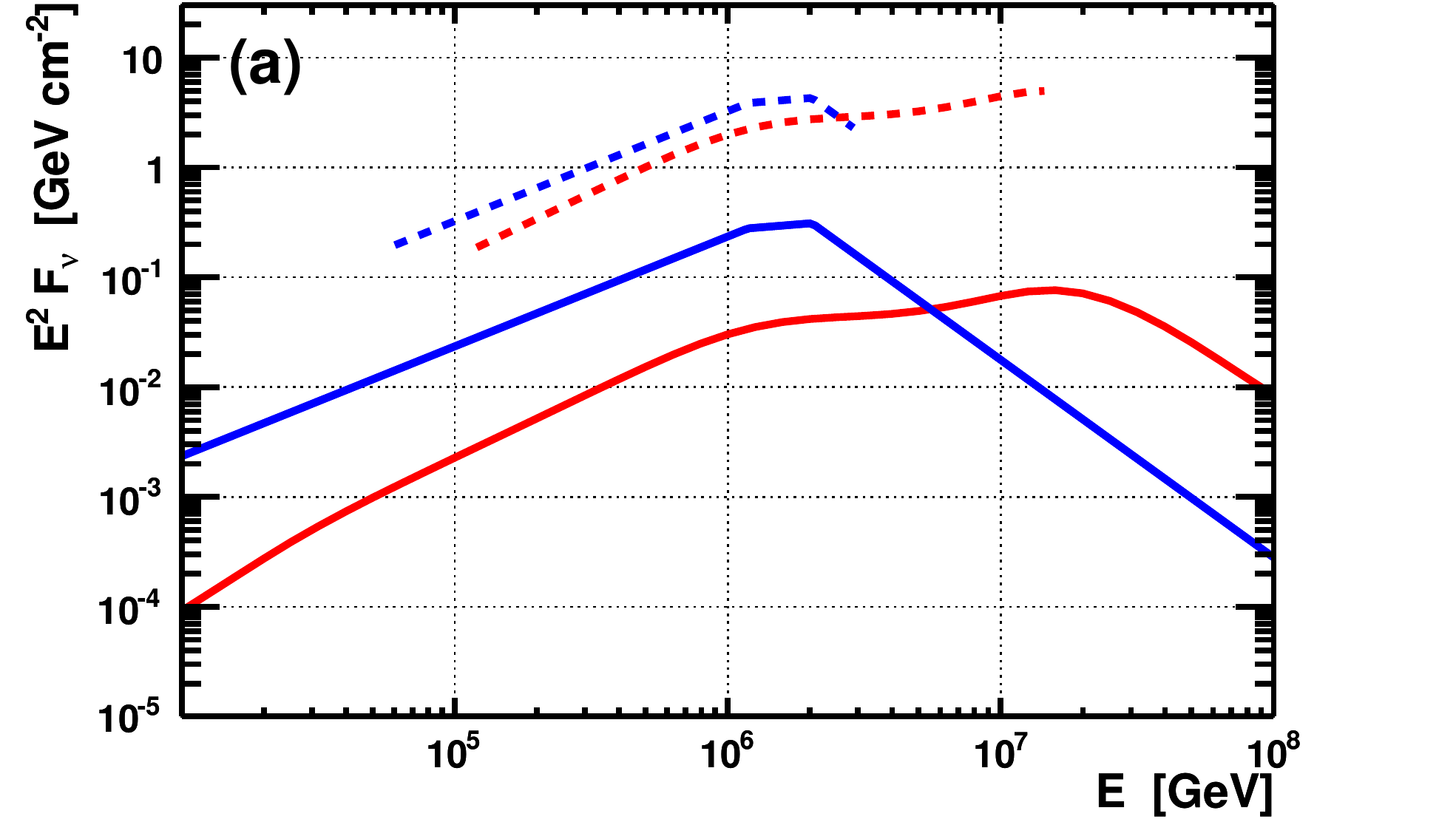}
  \includegraphics[width=\hsize]{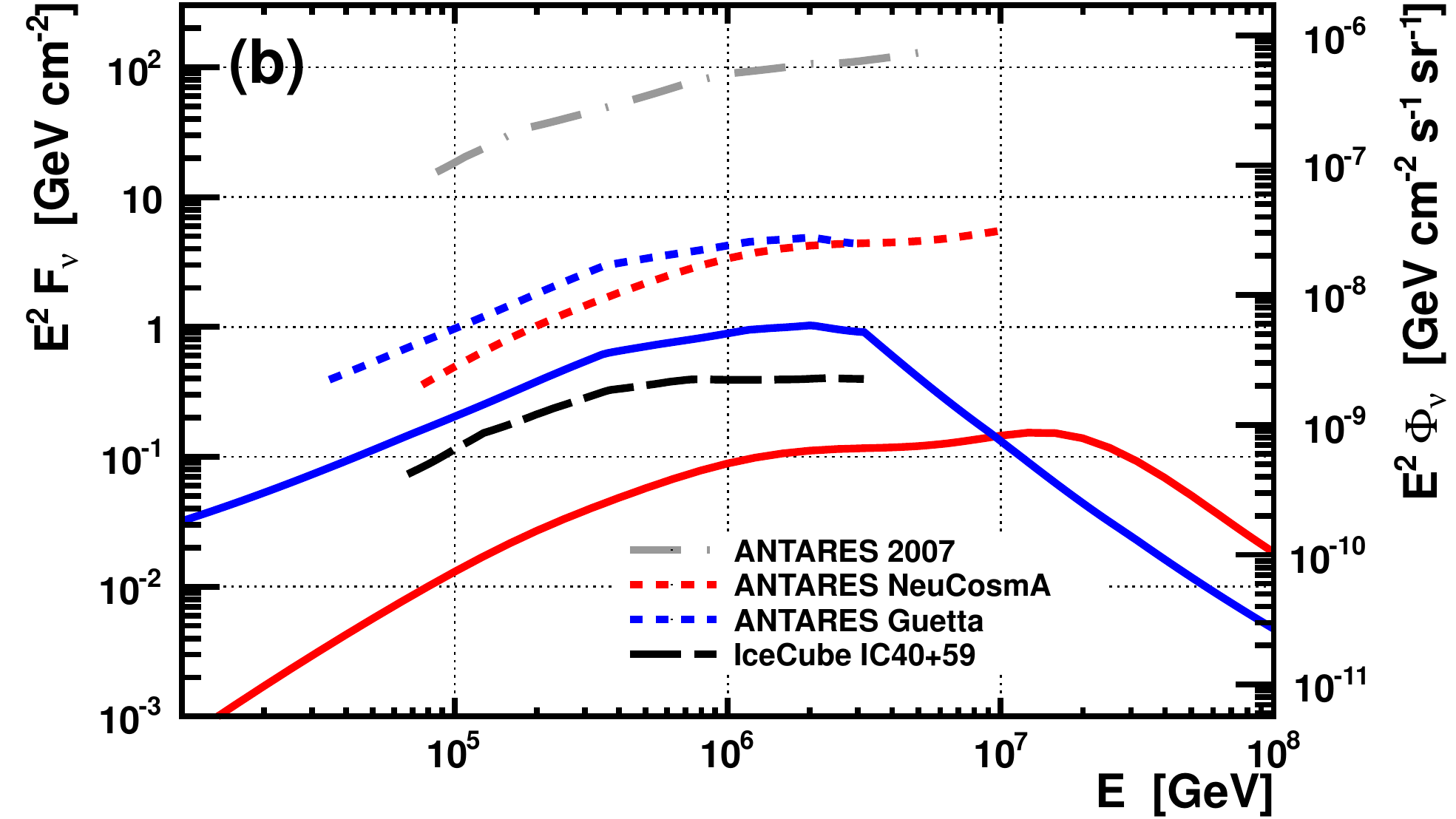}
  \caption{\label{fig:limit_antares} \label{fig:limit}
    (a): Expected muon neutrino spectra of the most promising burst GRB110918 (solid lines) from NeuCosmA \citep{Huemmer10a} (red) and \citet{Guetta04a} (blue). 
    Limits on these predictions are shown in the energy ranges where we expect 90\% of the flux (dashed lines).
    (b): Sum of the 296 individual gamma-ray-burst muon neutrino spectra (red and blue solid lines) and limits set by this analysis on the total flux expected from the sample (red and blue dashed lines). 
    The IceCube IC40+IC59 limit \citep[][]{IceCube12a} on the neutrino emission from 300 GRBs and the first ANTARES limit from 2007 using 40 GRBs \citep{Antares13a} are also shown in black (dashed) and grey (dash-dotted), respectively.
    The right-hand axis represents the inferred quasi-diffuse flux limit $E^2 \Phi_\nu$ (Eq.~\ref{eq:quasi_diffuse_flux}).
  }
\end{figure}

In total, 0.06 neutrino events from GRBs are expected from the NeuCosmA model, where only a small contribution of $4.6\cdot10^{-5}$ events is not due to particle tracks produced by muon neutrinos --
the Guetta model predicts 0.5 signals from muon neutrinos. 
The overall background in the $10\ensuremath{^\circ}$ cones is 0.05 events.
The 90\% C.L. upper limits on the expected number of  signal events $\mu_\mathrm{s}$ from each model are thus set to 2.3 events,
and the corresponding limits on the muon neutrino flux, $F_\nu$, from GRB110918 as well as on the cumulative flux from the whole sample are shown in Fig.~\ref{fig:limit}. The simple treatment of the Guetta model is represented here (see Fig.~\ref{fig:spectra} a, solid line).
For the NeuCosmA model, the limit on the total flux lies a factor of 38 above the expected spectrum (4.4 for Guetta).
The right-hand axis of  Fig.~\ref{fig:limit} (b) represents the limits translated into limits on the inferred quasi-diffuse neutrino flux: 
\begin{align}
  \label{eq:quasi_diffuse_flux}
  E^2 \Phi_\nu =  \sum E^2 F_\nu \times \frac{1}{4 \pi} \cdot\frac{1}{N_\mathrm{GRB}} \cdot 667\; \mathrm{y}^{-1} ,
\end{align}
where $\nu = \nu_\mu + \bar{\nu}_\mu$, assuming that each analysed sample represents an average burst distribution and that the annual rate of long bursts is 667 per year.

The first ANTARES limit \citep{Antares13a} obtained for 40 GRBs during the construction phase of the detector in the year 2007 is also shown in Fig.~\ref{fig:limit} (b). 
That analysis was based on the Guetta model (accounting for different break energies of $\nu_\mu$ and $\bar{\nu}_\mu$) and employed a  counting method searching for neutrino events in a two-degree cone around each burst. 
Using the data from the IC40 and IC59 detector phases in 2008 to 2010, IceCube recently published a more stringent limit on the neutrino emission as predicted by the `simple' Guetta model \citep{IceCube12a}, which is also shown in Fig.~\ref{fig:limit} (b).

Because of the larger effective area of the IceCube detector, the new ANTARES limit
presented in this paper does not set additional constraints on the Guetta emission model. 
Note, however, that both detectors have complementary sky coverage and therefore the analysed sample of GRBs differs significantly. 
90\% of the analysed bursts have not previously had their neutrino emission constrained.
When comparing limits obtained in different analyses, however, one should keep in mind that the precise shapes of the spectra -- and thus, of the limits
-- depend on the actual selected sample, the measured parameters of the individual bursts and their uncertainty, the set of default parameters and on the chosen model. 
\begin{table}
  \scriptsize
  \caption{Optimisation results for the ten most promising GRBs}
  \label{tab:analysis_params}
  \begin{tabular}{r| r | l l l | r | r | r} \hline \hline
    GRB & $\Lambda_\mathrm{cut}$ & $\mu_\mathrm{b}$ & $\mu_\mathrm{s}^\text{NeuCosmA}$ & $\mu_\mathrm{s}^\text{Guetta}$ & $\langle \alpha \rangle $ & $T_\mathrm{search}$ & $\sigma_\mathrm{tot}$ \\ 
    &  &  &  &  & (\ensuremath{^\circ}) & ($\mathrm{s}$) &\\
    \hline
    11091889  & -5.3 & 1.1$\cdot 10^{-4}$ & 3.0$\cdot 10^{-2}$ & 1.5$\cdot 10^{-1}$ & 0.30 & 73.4 & $\mathbf{5 \sigma}$ \\
    \hline
    11091889  & -5.5 & 3.7$\cdot 10^{-4}$ & 3.5$\cdot 10^{-2}$ & 1.7$\cdot 10^{-1}$ & 0.32 & 73.4  & \\ 
    08060725  & -5.4 & 5.5$\cdot 10^{-4}$ & 6.5$\cdot 10^{-3}$ & 1.4$\cdot 10^{-2}$ & 0.33 & 164.3  & \\ 
    11100892  & -5.5 & 3.6$\cdot 10^{-4}$ & 2.2$\cdot 10^{-3}$ & 2.6$\cdot 10^{-3}$ & 0.35 & 75.4  & \\ 
    10101417  & -5.1 & 4.1$\cdot 10^{-4}$ & 1.2$\cdot 10^{-3}$ & 1.7$\cdot 10^{-2}$ & 0.89 & 723.1  & \\ 
    10072809  & -5.6 & 2.0$\cdot 10^{-4}$ & 9.6$\cdot 10^{-4}$ & 1.4$\cdot 10^{-2}$ & 0.49 & 268.6  &\\ 
    09020174  & -5.4 & 5.4$\cdot 10^{-4}$ & 7.0$\cdot 10^{-4}$ & 2.4$\cdot 10^{-2}$ & 0.39 & 126.6  & \\ 
    11122048  & -5.2 & 1.4$\cdot 10^{-4}$ & 6.2$\cdot 10^{-4}$ & 1.2$\cdot 10^{-2}$ & 1.13 & 66.5  & \\ 
    09082967  & -5.4 & 1.7$\cdot 10^{-4}$ & 3.9$\cdot 10^{-4}$ & 5.7$\cdot 10^{-3}$ & 1.02 & 112.1  & \\ 
    11062215  & -5.4 & 1.7$\cdot 10^{-4}$ & 4.3$\cdot 10^{-4}$ & 9.5$\cdot 10^{-3}$ & 1.42 & 116.6  & \\ 
    08100914  & -5.5 & 1.3$\cdot 10^{-4}$ & 3.5$\cdot 10^{-4}$ & 1.9$\cdot 10^{-3}$ & 0.94 & 70.2  & \\ \cline{1-7} 
    all GRBs: &&&&&& &  $\mathbf{3\sigma}$ \\
    mean	& -5.4 & 1.7 $\cdot 10^{-4}$ & 2.0 $\cdot 10^{-4}$ & 1.6 $\cdot 10^{-3}$  & 2.85 & 80.4 \\
    sum 	& & 5.1$\cdot 10^{-2}$ & 6.1$\cdot 10^{-2}$ & 4.8$\cdot 10^{-1}$ & & 2.4$\cdot 10^{4}$& \\
    \hline 
  \end{tabular}
  \tablefoot{Optimised $\Lambda_\mathrm{cut}$ values for the ten gamma-ray bursts with the highest discovery probabilities and the resulting expected number of background and signal events $\mu_\mathrm{b}$ and $\mu_\mathrm{s}$ at the significance level $\sigma_\mathrm{tot}$. The consequent median angular spread of events $\langle \alpha \rangle$ is also provided. In the last rows, the sum and mean of the values for all 296 GRBs at $3\sigma$ is given. A full table is provided online\footnote{The full table is available in electronic form at the CDS via anonymous ftp to cdsarc.u-strasbg.fr (130.79.128.5) or via \url{http://cdsarc.u-strasbg.fr/viz-bin/qcat?J/A+A/559/A9}}. The naming convention for the GRBs is similar to that used by \textsl{Fermi}, the last two digits of the GRB name correspond to the fraction of the day at which the burst occurred.}
\end{table} 

\section{Conclusion}
\label{sec:discussion}
Using data from the ANTARES detector,  a search for muon neutrinos in coincidence with 296 gamma-ray bursts occurring between the end of 2007 and 2011 has been performed.  
No events passed the selection criteria and limits on the neutrino flux were derived. 
For the NeuCosmA model, a limit on $E^2 F_\nu$ of $0.35 - 5.6 \; \mathrm{GeV}\; \mathrm{cm}^{-2}$ in the energy range from $7.5\cdot10^{4}\; \mathrm{GeV}$ to $1.0\cdot10^{7}\; \mathrm{GeV}$ was derived and compared with limits obtained in previous analyses.

This work is the first analysis based on an advanced numerical calculation of GRB neutrinos:
the NeuCosmA code includes full photohadronic interaction cross-sections, energy losses of secondary particles, and flavour mixing. 
The neutrino flux has been shown to be an order of magnitude below that predicted by previous analytic approaches. 
This helps to resolve the tension between the non-observation of a neutrino signal and the most stringent experimental constraint currently available \citep{IceCube12a}, which was a factor of 2.1 below the predictions made by the Guetta model. 

Hence, existing limits do not yet constrain realistic neutrino emission models based on an internal shock scenario. 
Nevertheless, the collection of more data with active experiments such as ANTARES and IceCube, as well as with the planned neutrino telescope KM3NeT, will certainly allow the widely established fireball paradigm for gamma-ray bursts to be probed in the near future.

\begin{acknowledgements}
The authors acknowledge the financial support of the funding agencies:
Centre National de la Recherche Scientifique (CNRS), Commissariat \`a
l'\'ene\-gie atomique et aux \'energies alternatives (CEA), Agence
National de la Recherche (ANR), Commission Europ\'eenne (FEDER fund
and Marie Curie Program), R\'egion Alsace (contrat CPER), R\'egion
Provence-Alpes-C\^ote d'Azur, D\'e\-par\-tement du Var and Ville de La
Seyne-sur-Mer, France; Bun\-des\-mi\-nis\-te\-ri\-um f\"ur Bildung und Forschung
(BMBF), Germany; Istituto Nazionale di Fisica Nucleare (INFN), Italy;
Stichting voor Fundamenteel Onderzoek der Materie (FOM), Nederlandse
organisatie voor Wetenschappelijk Onderzoek (NWO), the Netherlands;
Council of the President of the Russian Federation for young
scientists and leading scientific schools supporting grants, Russia;
National Authority for Scientific Research (ANCS), Romania; Ministerio
de Ciencia e Innovaci\'on (MICINN), Prometeo of Generalitat Valenciana
and MultiDark, Spain; Agence de l'Oriental and CNRST, Morocco. We also
acknowledge the technical support of Ifremer, AIM and Foselev Marine
for the sea operation and the CC-IN2P3 for the computing facilities.
We would like to thank Walter Winter for helpful discussions and making it possible to use the NeuCosmA model.
\end{acknowledgements}

\appendix

\section{GRB selection}
\label{app:grb_selection}
The table of the \textsl{Swift} satellite\footnote{Swift: \url{http://swift.gsfc.nasa.gov/docs/swift/archive/grb_table.html}} contains data from the three on-board instruments BAT (gamma rays), XRT (X-rays), and UVOT (ultraviolet), ordered with increasing position-measurement accuracy from arcminutes to sub-arcseconds. 
The information provides BAT spectral measurements in the energy range from 15 to $150\; \mathrm{keV}$. 
The \textsl{Swift} BAT2 Catalogue \citep{Sakamoto11a}\footnote{BAT2: \url{http://vizier.u-strasbg.fr/viz-bin/VizieR?-source=J/ApJS/195/2}} provides re-analysed \textsl{Swift} data, so the spectral information therein is considered to be more accurate. 
The \textsl{Fermi} GBM Burst Catalogue\footnote{Fermi: \url{http://heasarc.gsfc.nasa.gov/W3Browse/fermi/fermigbrst.html}} \citep{Goldstein12a,Paciesas12a} supplies the best spectral information in the energy range from $10\; \mathrm{keV}$ to $1\; \mathrm{MeV}$: 
the peak flux spectrum and the spectrum averaged over the burst duration (which is eventually used for the neutrino spectrum calculation) is fitted with four different spectral functions. 
The angular resolution is of the order of degrees. 
The IceCube Collaboration also provides a table with GRB parameters\footnote{IceCube: \url{http:// grbweb.icecube.wisc.edu}} \citep{IceCube11a}, which is created by parsing the Gamma-ray Coordinates Network (GCN) notices\footnote{GCN: \url{http://gcn.gsfc.nasa.gov/gcn3_archive.html}}. 
This table is used to fill up missing parameter values for GRBs that have been found in at least one of the other tables. 

When merging the information on the gamma-ray burst parameters, we assigned priorities to the measured values according to their considered accuracy. 
The priorities of parameters are shown in Table~\ref{tab:tables_priorities} in square brackets, as well as the percentage of how often information was obtained from each source.
\begin{table*}
\caption{Usage of the GRB parameter catalogues.}
\centering 
\label{tab:tables_priorities}
\begin{tabular}{l rrrrrrr} \hline \hline 
Source & Position & Time & Fluence & Spectrum & Duration & Redshift & Start \& Stop
\\ \hline 
Swift:BAT & 3.7\% \text{[3]} & 11.2\%   \text{[3]}&  10.5\% \text{[3]}&  14.2\% \text{[3]}& 8.8\% \text{[3]} &  4.4\% \text{[2]} & \; \; ---  \\
Swift:XRT &17.2\%  \text{[2]} &&&&&&\\
Swift:UVOT &  11.2\%  \text{[1]}& & & &&&\\ \hline
Swift BAT2 & &   11.2\% \text{[2]}&  10.1\% \text{[2]}& 9.8 \% \text{[2]}  &11.2\% \text{[2]}  &   4.4 \%  \text{[1]} &  10.1\% \text{[2]}\\ \hline
Fermi & 67.9\% \text{[4]}&  77.7\%  \text{[1]}  & 77.7\% \text{[1]}&  36.2\% \text{[1]} &   77.4\% \text{[1]}&  --- &   77.7\% \text{[1]}\\ \hline
IceCube  & &  &   1.7\% \text{[4]}& 4.4\% \text{[4]}&2.4\% \text{[4]} &  0.3\% \text{[3]}& 11.2\% \text{[3]} \\
\hline 
\end{tabular}
\tablefoot{The numbers in square brackets give the assigned priority of this source of information with respect to the parameter(s).}
\end{table*}

\label{sec:default_values}
When a parameter could not be measured, standard values as given in Table~\ref{tab:default_values} were used to calculate the spectra. 
The form of the photon spectrum is determined by the spectral indices $\alpha$ and $\beta$ with the break energy $\epsilon_\mathrm{peak}$ giving their transition. The isotropic luminosity $L_\mathrm{iso}$ can be calculated from the redshift $z$ and the total measured fluence in gamma rays $\mathcal{F}$ (given in the energy range from $E_\mathrm{min}$ to $E_\mathrm{max}$) via $L_\mathrm{iso}=4 \pi d_\mathrm{L}^2 \frac{\mathcal{F}}{T_{90}}$ with the luminosity distance $d_\mathrm{L}$. 
In case of unknown redshift $z$, we took the default value of $L_\mathrm{iso}$.
$T_{90}$ is the time in which 90\% of the fluence is emitted. 
The other parameters such as the jet Lorentz boost factor $\Gamma$, the fraction of jet energy in electrons $\epsilon_e$ and in the magnetic field $\epsilon_B$, the ratio of energy in electrons and protons $f_e$, the average fraction of proton energy transferred to a pion $\langle x_{p\rightarrow \pi} \rangle$ and the variability of the gamma-ray light curve $t_\mathrm{var}$ 
are not present in the tables and hence were taken as default.
The standard values are the same as given in \citet{IceCube11a}, with some differences to the IC22 \citep{IceCube10a} default values: $z=2.0$, $\Gamma=300$, $L_\mathrm{iso}=10^{51}\; \mathrm{erg} \; \mathrm{s}^{-1}$.
\citet{Baerwald12b} give a very elaborate overview about the NeuCosmA spectra changing with the input parameters.

The parameters of the extraordinarily strong GRB110918 are presented in Table~\ref{tab:params_110918}. 

The time window of the search, $T_\mathrm{search}$, for emission from each burst is delineated by the start and stop times as measured by the satellites or, when these are not provided in the catalogues, as  $T_{90} \pm 30\%$.
Additionally, we accounted for the detector's data acquisition uncertainty ($0.4\; \mathrm{s}$), the satellite time given in integer seconds ($1\; \mathrm{s}$), and the light propagation from a satellite through Earth to the detector ($ 0.5\; \mathrm{s}$) by adding another $\pm 2\; \mathrm{s}$ to the search-time window.

\begin{table}
  \caption{Standard gamma-ray-burst parameters as described in the text.}\label{tab:default_values}
  \small
  \begin{tabular}{lll} \hline \hline 
    $\alpha = 1$ 	&    $\beta =  \alpha +1$ 	&	$\epsilon_\mathrm{peak} = 200\; \mathrm{keV}$  \\
    $z = 2.15$ &  $L_\mathrm{iso}  =10^{52}\; \mathrm{erg} \; \mathrm{s}^{-1} $  &  \\
    $\Gamma = 316$	& $\epsilon_e = 0.1 $& $\epsilon_B = 0.1$ \\
    $f_e = 0.1$  &$\langle x_{p\rightarrow \pi} \rangle = 0.2$ & $t_\mathrm{var} = 0.01\; \mathrm{s}$ \\
    \hline 
  \end{tabular}
\end{table}

\begin{table}
  \caption{Gamma-ray-burst parameters of GRB110918 as described in the text.}\label{tab:params_110918}
  \small
  \begin{tabular}{lll} \hline \hline 
    $\alpha		= 1.2 				$	& $\beta 	= 2.0 	$	& $\epsilon_\mathrm{peak} = 150 \; \mathrm{keV} $\\ 
    $\mathcal{F}	= 7.5 \cdot 10^{-4}\; \mathrm{erg} \; \mathrm{cm}^{-2} $	& $E_\mathrm{min} 	= 0.02 \; \mathrm{MeV} $	& $E_\mathrm{max} = 10 \; \mathrm{MeV}  $ \\
    $\text{UT} = 21:26:57$ & \textsl{dec}$= 32.5\ensuremath{^\circ} $\tablefootmark{a} & \textsl{RA}$= -27.1\ensuremath{^\circ} $\tablefootmark{a} \\
    $\Delta_\mathrm{err} = 0.5\arcsec$ & $ T_{100} 	= 69.4 \; \mathrm{s}$\tablefootmark{b}		&$ z 		= 0.982$ \tablefootmark{c} \\
    \hline 
  \end{tabular}
  \tablefoot{All values are read from the IceCube table. Values measured by \textsl{Konus-Wind} \citep{GCN12362}, if not marked otherwise.
    \tablefoottext{a}{Measured by the \textsl{Isaac Newton Telescope} \citep{GCN12365}.}
    \tablefoottext{b}{Integration time of \textsl{Konus-Wind} \citep{GCN12362}.}
    \tablefoottext{c}{Determined from \textsl{Gemini-N} \citep{GCN12368} and the \textsl{GTC} telescope \citep{GCN12375}.}
  }
\end{table}

\section{Background estimation} \label{app:background}
We estimated the background event rate for each GRB separately.
First, the time-averaged reconstructed event rate in the data from late 2007 to 2011 from the direction of the GRB was estimated. 
Either the rate averaged over all data-taking runs at the GRB's position $(\theta, \phi)_\mathrm{GRB} $ was used, or -- if resulting in a higher rate --  the mean of the corresponding time-averaged rates within a $10\ensuremath{^\circ}$ cone around this position. This establishes a conservative background estimate, accounting for non-uniformity of the background in the vicinity of the GRB's position.

To take into account the varying efficiency of the detector with time, this average rate was then scaled by a correction factor  $c_i$ for each data-taking run $i$ of $\sim 2.5$ hours. 
Each $c_i$ was calculated by the ratio of the total number of events (in all directions) in the corresponding run $n_i$ to the average total number of events for the respective run duration $t_i$ (see Eq.~\ref{eq:background}).
As $n_i$ may be very small for short runs, the 90\% C.L. upper limit was used instead.
Additionally, factors for specific run periods $c_\mathrm{period}$ were applied taking into account differences between longer phases of similar run conditions. These values were obtained by fitting the background rate in certain periods separately. 
This approach assumes that the total number of events -- dominated mostly by downgoing atmospheric muons -- is proportional to the number of upgoing events. 
To test this assumption, we determined the measured and estimated rates of upgoing events in longer time periods of a few days, excluding data-taking runs in which GRBs occurred. 
The measured rate was always found to be  $\mu_\mathrm{meas} <1.5 \cdot \mu_\mathrm{est}$, thus we conservatively increased the estimate by 50\%.
Consequently, the expected number of background events in coincidence with each gamma-ray burst search-time window $T_\mathrm{search}$ was calculated via
\begin{equation}
  \begin{aligned}
    \mu_\mathrm{b} (\theta, \phi)_\mathrm{GRB} &=  T_\mathrm{search} \times
    \langle  \mu (\theta, \phi)_\mathrm{GRB}  \rangle_\text{all runs} \cdot c_i  
    \cdot c_\mathrm{period} \cdot 1.5 \\
    \text{with  } c_i &= \frac{\left[ n_i \right]^{90\%}}{t_i \sum n_j / \sum t_j}  \, \text{, where $j$ includes all data-taking runs.}
  \end{aligned} \label{eq:background}
\end{equation}

\bibliographystyle{./aa-package/bibtex/aa}
\bibliography{./mnemonic,./aa_abbrv,./bibfile}	

\end{document}